\documentclass[sigconf]{acmart}
\setcopyright{none}
\usepackage{algorithm}
\usepackage{algpseudocode}
\usepackage{adjustbox}
\usepackage[capitalize,noabbrev]{cleveref}
\usepackage{amsmath}
\usepackage{microtype}
\usepackage{float}
\usepackage{multirow}
\theoremstyle{plain}
\newtheorem{theorem}{Theorem}[section]
\newtheorem{proposition}[theorem]{Proposition}
\newtheorem{lemma}[theorem]{Lemma}
\newtheorem{corollary}[theorem]{Corollary}
\theoremstyle{definition}
\newtheorem{definition}[theorem]{Definition}

\theoremstyle{remark}
\newtheorem{remark}[theorem]{Remark}

\usepackage{graphicx}
\usepackage{textcomp}
\usepackage{xcolor}

\newcommand{\revision}[1]{\textcolor{black}{#1}}

\begin{document}

\title{Beyond Right to be Forgotten: Managing Heterogeneity Side Effects Through Strategic Incentives}

\author{Jiaqi Shao$^{1}$, Tao Lin$^{2}$, Xiaojin Zhang$^{3}$, Qiang Yang$^{4,5}$, Bing Luo$^{1}$}
\renewcommand{\shortauthors}{Shao et al.}

\affiliation{$^{1}$Duke Kunshan University, Suzhou\country{China}}
  
% }
\affiliation{$^{2}$Westlake University, Hangzhou\country{China}
}
\affiliation{$^3$ Huazhong University of Science and Technology, Wuhan\country{China}}

\affiliation{$^4$Hong Kong University of Science and Technology, Hong Kong\country{China}
}
\affiliation{$^{5}$WeBank, Shenzhen\country{China}
}

% \email{xiaojinzhang@hust.edu.cn}
% \email{qyang@cs.ust.hk}
% \email{bing.luo@dukekunshan.edu.cn}

% Define a shorter version of the authors' names for the page headers
\renewcommand{\shortauthors}{Shao et al.}

\begin{abstract}
    Federated Unlearning (FU) enables the removal of specific clients' data influence from trained models. 
    However, in non-IID settings, removing clients creates critical side effects: 
    remaining clients with similar data distributions suffer disproportionate performance degradation, 
    while the global model's stability deteriorates. 
    These vulnerable clients then have reduced incentives to stay in the federation, 
    potentially triggering a cascade of withdrawals that further destabilize the system.
    To address this challenge, we develop a theoretical framework that quantifies how data heterogeneity impacts unlearning outcomes.
    Based on these insights, we model FU as a Stackelberg game where the server strategically offers payments to retain crucial clients 
    based on their contribution to both unlearning effectiveness and system stability.
    Our rigorous equilibrium analysis reveals how data heterogeneity fundamentally shapes the trade-offs between system-wide objectives and client interests.
    Our approach improves global stability by up to 6.23\%, 
    reduces worst-case client degradation by 10.05\%, 
    and achieves up to 38.6\% runtime efficiency over complete retraining.
\end{abstract}

    


\maketitle
\section{Introduction}

\begin{figure}[t]
    \centering    \includegraphics[width=0.9\linewidth]{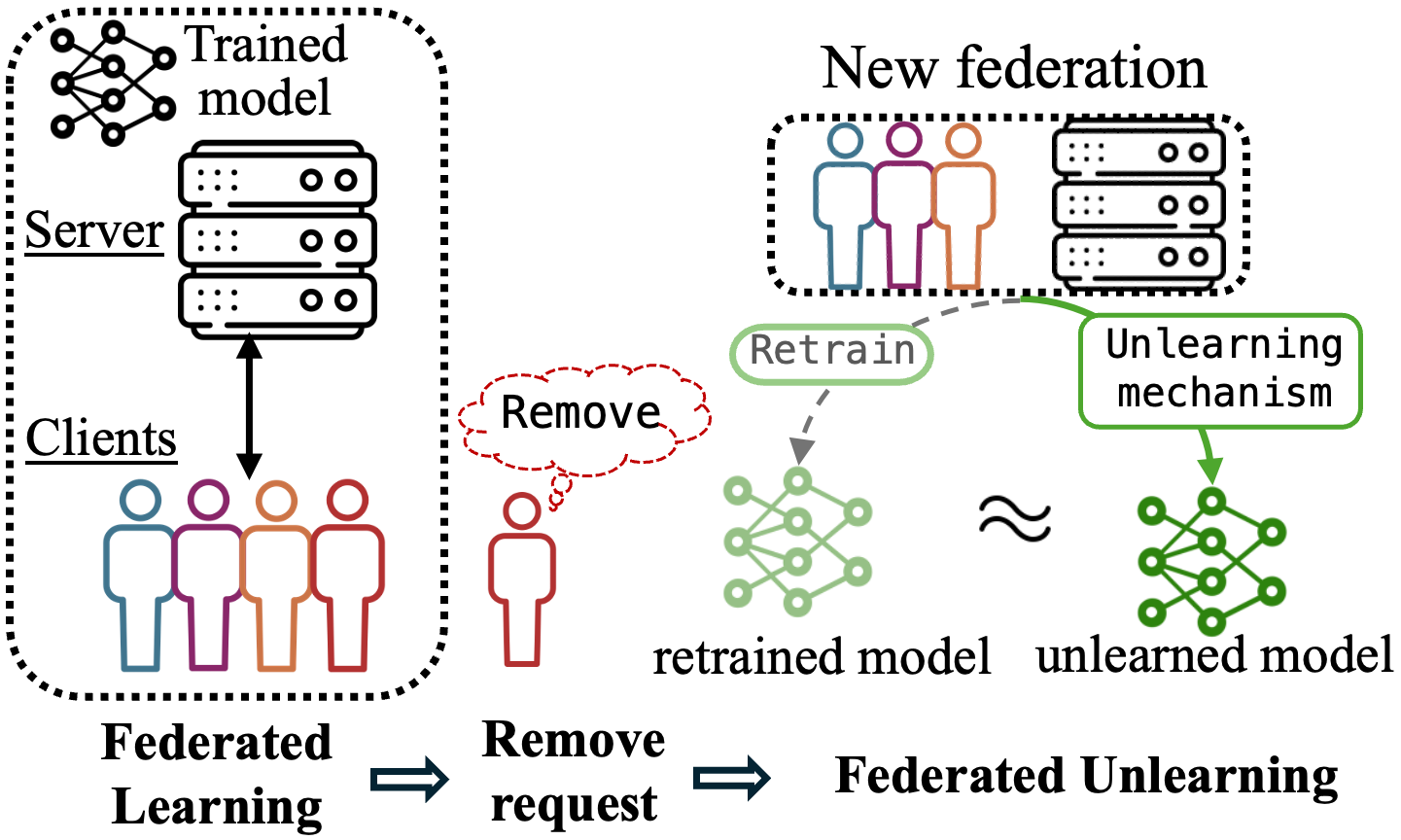}
    \caption{An illustration of the Federated Unlearning~(FU) process: Begin with a client's removal request, leading to a new federation with remaining clients. The FU mechanism then generates an unlearned model that approximates a model retrained from scratch. }
    \Description{An illustration of the Federated Unlearning~(FU) process: Begin with a client's removal request, leading to a new federation with remaining clients. The FU mechanism then generates an unlearned model that approximates a model retrained from scratch.}
    \label{fig:model}
    % \vspace{-2em}
\end{figure}

Federated learning (FL) is a promising paradigm for collaborative machine learning, allowing multiple data owners to train a shared model without exchanging raw data~\cite{mcmahan2017communication, tran2019federated,wang2019beyond}. However, as data privacy regulations (such as GDPR~\cite{regulation2018general} and CCPA~\cite{goldman2020introduction}) become more stringent, individuals grow increasingly aware of their digital rights, including ``the right to be forgotten"~\cite{liu2021federaser}.
In response to these evolving privacy concerns and compliance requirements, Federated Unlearning~(FU) has emerged to allow the removal of the influence of a client from a trained model~\cite{liu2021federaser,wang2023federated,jeong2024sok}, illustrated in~\cref{fig:model}. 
% FU complements the traditional FL process by providing a mechanism to remove the impact of specific clients' data from the global model. This integrated FL-FU pipeline ensures that the system allows clients to withdraw their contributions when requested.

In FU, the intuitive approach to removing a client's influence is to retrain the model from scratch, excluding the data of the removed clients. 
However, this method is computationally expensive, especially for large-scale models and datasets~\cite{liu2022right}. 
Recent works in FU primarily focus on efficiently achieving \textit{unlearning effectiveness} that approximates complete retraining through various techniques, such as leveraging remaining client data or utilizing server-side historical information to produce an unlearned model~(\textit{e}.\textit{g}.,~\cite{liu2021federaser, fraboni2022sequential, gao_verifi_2022,su2023asynchronous,wuerkaixi2024accurate,lin2024incentive}).
While these approaches have advanced efficiency, they overlook a critical vulnerability in federated systems: \textbf{the inherent data heterogeneity across clients creates systemic side effects during unlearning}, which introduce:\looseness=-1

1. \textit{Local performance degradation:} Removing clients can disproportionately impact specific remaining clients whose data distributions share similarities with the removed clients, with our experiments revealing up to 13.34\% degradation for affected clients, threatening the collaborative foundation of federated systems.\looseness=-1

2. \textit{Global stability deterioration:} When distinctive distributional components are removed, the model's generalization capabilities can deteriorate, leading to jeopardized system reliability.

In response to these challenges, 
we identify the strategic participation problem at the core of these side effects. 
When clients are removed, those with similar data distributions 
suffer disproportionate performance degradation. 
These vulnerable clients then have reduced incentives to remain in the federation, 
potentially triggering a cascade of withdrawals that further destabilize the system.
Our key insight is that 
an incentive mechanism can strategically retain crucial clients by compensating them according 
to their contribution to both unlearning effectiveness and system stability. 

However, modeling the complex interactions between data heterogeneity and unlearning outcomes presents significant challenges.
The multi-dimensional impact of client distribution shifts during unlearning creates analytical difficulties in quantifying the side effects and unlearning effectiveness. 
Thus, this leads to our first key research question:

\noindent\textit{\textbf{RQ1}: How to rigorously model the complex relationships between data heterogeneity, unlearning effectiveness, and the multi-dimensional side effects in Federated Unlearning?}

The intricate relationships between non-IID data distributions and their impact on unlearning outcomes remain unexplored, 
making it impossible to quantify how removing clients affects both individual client performance and global model stability in FU. 
We first develop a theoretical framework based on kernel mean embeddings that quantify heterogeneity-induced side effects. 
With this heterogeneity-aware framework, we can now leverage these insights to study the second key research question:

\noindent\textit{\textbf{RQ2}: How to design a robust incentive mechanism that 
strategically balances objectives between the server and clients while addressing heterogeneity-induced side effects in FU?}

In FU, the server's objective (unlearning effectiveness and stability) and the client's objective (mitigation of performance degradation) 
create complex strategic dynamics, which are further complicated by clients' strategic behaviors, 
where each client's participation decision 
 affects others' outcomes through data distribution shifts. 
To address this challenge, we develop a novel incentive mechanism that 
mitigates heterogeneity-induced side effects in FU through a bi-level optimization framework. 
We also derive closed-form expressions for utility functions and strategic interactions, 
providing a comprehensive equilibrium guarantee. 

Our key contributions are summarized as follows:

$\bullet$ \textit{We are the first to identify, quantify, and address the heterogeneity-induced side effects in FU}, advancing beyond mere compliance with the right to be forgotten. 
Our novel incentive mechanism resolves the fundamental trade-off between system-wide objectives (unlearning effectiveness and global stability) and unlearning impacts across clients.

$\bullet$ \textit{We establish a rigorous analytical foundation to quantify the multi-dimensional impact of unlearning} on both heterogeneity-induced side effects in FU. 
To capture the complex interplay between heterogeneity and unlearning outcomes, we leverage kernel mean embeddings (KME) to develop heterogeneity-aware performance bounds that enable tractable analysis of strategic client behavior under heterogeneous data conditions.

$\bullet$ \textit{We formulate the FU incentive design problem as a two-stage Stackelberg game and prove the existence of the Nash equilibrium and its uniqueness} among heterogeneous clients despite complex interdependencies. 
We further characterize optimal server strategies through closed-form expressions
revealing how data heterogeneity fundamentally shapes the unlearning outcomes.
Our algorithmic contribution transforms the nonconvex optimization into 
a computationally efficient quasiconvex formulation 
with theoretical convergence guarantees.

$\bullet$ 
We evaluate our approach on BERT-AG\_NEWS (text classification) and ResNet-CIFAR100 (image classification) under 
varying degrees of data heterogeneity.
 Our method consistently outperforms baselines, 
 improving global stability by 6.23\% and 
 reducing worst-case client performance degradation by 10.05\% 
 while maintaining unlearning effectiveness. 
 In addition, our method achieves up to 38.6\% runtime efficiency over complete retraining.

% Due to space constraints, we leave the detailed theoretical proof to the supplementary material~\cite{shao2025supplementary}.
Due to space constraints, we leave the detailed theoretical proof to the supplementary material~(\cref{app:lemma:loss_kernel_mean}-\ref{app:mono-px}).

\section{Related Work \& Preliminaries}
\subsection{Related Work}

Existing FU techniques primarily focus on effectively removing the influence of the target data. These approaches can be broadly categorized into methods using historical information~\cite{liu2021federaser,wang2023mitigating,fraboni2024sifu}, gradient manipulation~\cite{zhao2023federated,gao2024verifi}, knowledge distillation~\cite{zhu2023heterogeneous}, continuous learning~\cite{shibata2021learning,liu2022continual,shaik2024framu,shao2024federated} and reverse training~\cite{dhasade2023QuickDrop,li2023subspace}. 
While Shao et al.~\cite{shao2024federated} have investigated trade-offs between stability/fairness and unlearning effectiveness, a comprehensive framework that \textit{simultaneously} addresses system-wide (server-side) FU objectives with client-side concerns remains underexplored.

Recent studies on FU have explored incentive design for both data removal requests and the unlearning process~\cite{ding2024strategic,ding2023incentive}. These studies examine incentives for data revocation, aiming to retain users while balancing partial revocation with model performance. 
However, the model performance evaluation in these works relies on the assumption of diminishing marginal utility for additional data, \textit{overlooking the critical role of data heterogeneity}.\looseness=-1

Additionally, some works focus on incentivizing remaining clients to participate in the unlearning process given removal requests. 
Lin et al.~\cite{lin2024incentive} proposed an incentive mechanism considering remaining clients' utility in terms of payment and participation cost for additional unlearning computations. Building on this,~\citet{liu2024guaranteeing} incorporated privacy considerations by integrating secure aggregation techniques into the incentive design.
However, they overlooked the heterogeneous impact of unlearning on individual clients and focused on either client utility or server objectives without capturing their complex interplay. \looseness=-1

While incentive design for client sampling in FL has been extensively studied (\textit{e}.\textit{g}.,~\cite{deng2021fair,tang2021incentive,luo2022tackling,ding2023joint,luo2023incentive,li2024adafl}), these methods primarily address bias reduction for a static data distribution. 
\revision{In contrast, FU introduces fundamentally different challenges due to the dynamic nature of data distributions following client removal, 
requiring approaches that simultaneously balance unlearning effectiveness with system stability and individual client performance concerns.}

To address these limitations, our work focuses on incentivizing remaining clients to participate in unlearning. 
We incorporate two crucial factors overlooked in previous studies: 
(1) Explicit modeling of heterogeneity in FU evaluations, where heterogeneity reflects the dynamic nature of client data distributions and their varying impacts on unlearning effectiveness. 
(2) Simultaneous consideration of multiple objectives in FU, including unlearning effectiveness, global model stability, and client-side performance. 

\subsection{Preliminaries}

In this section, we provide an overview of federated learning and unlearning, introducing the formulation of FL, the objectives of FU, and the metrics used to evaluate unlearning effectiveness, global stability, and client performance change.
\subsubsection{Federated learning (FL)}
FL is a distributed machine learning paradigm that enables training models on decentralized data~\cite{mcmahan2017communication}. 

Let $\mathcal{O} = \{1,\ldots,M\}$ be a set of clients, where each client $i$ has $n_i$ local data points with data distribution $\mathcal{D}_i$.
    The optimal model $\boldsymbol{w}^o \in \mathcal{W}$ in FL is typically obtained by~\cite{mcmahan2017communication}:
\begin{equation}\label{eq:fl_obj}
\boldsymbol{w}^o = \text{argmin}_{\boldsymbol{w} \in \mathcal{W}} \left[F(\boldsymbol{w}) := \sum_{i \in \mathcal{O}} \frac{n_i}{n_\mathcal{O}} f_i(\boldsymbol{w})\right] \,,
\end{equation}
where $n_\mathcal{O} = \sum_{i \in \mathcal{O}} n_i$, and $f_i(\boldsymbol{w}) = \mathbb{E}_{z \sim \mathcal{D}_i}[\ell(\boldsymbol{w}, z)]$ is the empirical local loss for client $i$ with the loss function $\ell(\cdot)$.

\subsubsection{Federated Unlearning (FU)}
FU is a mechanism to remove the influence of target clients from a trained federated model. 
Let $\mathcal{R} \subset \mathcal{O}$ be the set of clients requesting data removal, and let $\mathcal{N} = \mathcal{O} \setminus \mathcal{R}$ be the set of remaining clients. The goal of FU is to obtain a model that minimizes the global loss over the remaining clients~\cite{ding2024strategic}:
\begin{equation}\label{eq:fu_obj}
\boldsymbol{w}^r = \text{argmin}_{\boldsymbol{w} \in \mathcal{W}} \left[F_\mathcal{N}(\boldsymbol{w}) := \sum_{i \in \mathcal{N}} \alpha_i f_i(\boldsymbol{w})\right] \,,
\end{equation}
where $\alpha_i = \frac{n_i}{n_\mathcal{N}}$, with $n_\mathcal{N} = \sum_{i \in \mathcal{N}} n_i$ representing the total number of data points across the remaining clients. 
In practice, the optimal unlearned model $\boldsymbol{w}^r$ can be obtained through \textit{complete retraining} on the remaining clients, but this process is often computationally expensive.

To evaluate the effectiveness of unlearning and its impact on both the system and individual clients, we utilize the following metrics adapted from~\cite{shao2024federated}. 
These metrics compare the performance of the unlearned model $\boldsymbol{w}^u$ (obtained through an unlearning mechanism) against the optimal unlearned model $\boldsymbol{w}^r$, assessing how well the unlearning outcomes approximate complete retraining:

\begin{definition}[Unlearning Effectiveness]\label{def:fu_effective}
The unlearning effectiveness metric $V$ measures how well an unlearned model approximates the optimal model that would be obtained by complete retraining on the remaining clients:
\begin{equation}\label{eq:fu_effective}
V(\boldsymbol{w}) = F_\mathcal{N}(\boldsymbol{w}) - F_\mathcal{N}(\boldsymbol{w}^r) \,,
\end{equation}
where $F_\mathcal{N}(\cdot)$ is the global loss function over the \textit{remaining clients} $\mathcal{N}$ and $\boldsymbol{w}^r$ is the optimal retrained model as defined in~\eqref{eq:fu_obj}. A smaller value of $V(\boldsymbol{w})$ indicates better unlearning performance, with $V(\boldsymbol{w}) = 0$ achieved when $\boldsymbol{w}$ perfectly matches the retrained model.
\end{definition}
\begin{definition}[Global Stability]\label{def:fu_stab}
The global stability metric $S$ measures the change in the overall performance of the unlearned model relative to the original FL system:
\begin{equation}\label{eq:fu_stab}
S(\boldsymbol{w}) = F(\boldsymbol{w}) - F(\boldsymbol{w}^o) \,,
\end{equation}
where $F(\cdot)$ is the global loss function over all clients $\mathcal{O}$, and $\boldsymbol{w}^o$ is the original FL-trained model before unlearning, 
as defined in~\eqref{eq:fl_obj}.
 A smaller value of $S(\boldsymbol{w}^u)$ compared to $S(\boldsymbol{w}^r)$ 
 indicates that the unlearned model better preserves the system's overall performance compared to the retrained model.
 \revision{We evaluate global stability on the full original set $\mathcal{O}$ to assess broader systemic effects of unlearning, aligning with our goal to go beyond ``the right to be forgotten" and examine its impact on overall model generalization.}
\end{definition}

\begin{definition}[Client Performance Change]\label{def:fu_local}
For each remaining client $i \in \mathcal{N}$, the performance change metric $Q_i$ measures the change in the client's local loss after unlearning:
\begin{equation}\label{eq:fu_local}
Q_i(\boldsymbol{w}) = f_i(\boldsymbol{w}) - f_i(\boldsymbol{w}^o) \,,
\end{equation}
where $f_i(\cdot)$ is the local loss function for client $i$, and $\boldsymbol{w}^o$ is the original FL-trained model before unlearning as defined in~\eqref{eq:fl_obj}.  A positive value of $Q_i(\boldsymbol{w})$ indicates performance degradation for client $i$ after unlearning, while a negative value represents performance improvement.
\end{definition}
% Ideally, we aim to minimize $Q_i$ for all clients to ensure fairness.

\section{System Model and Problem Formulation}

\revision{In this section, we outline our FU framework. 
When a client requests data removal, 
the system transitions to the unlearning phase that introduces ``dynamic client participation''. }

Specifically, during the FU process, when certain clients request to be removed from a trained federated model, the server needs to collaborate with the remaining clients $\mathcal{N} = \{1, \dots, N\}$ to effectively eliminate the influence of the removed data. 
This creates a strategic interaction between two parties: (a) the \textit{server}, which balances unlearning effectiveness with system stability to prevent model performance fluctuations; and (b) the \textit{remaining clients}, who weigh participation costs against potential local performance degradation, especially when removed clients share similar distributions with their own. 
  
In the following, we define utility functions and decision problems for the server and clients, considering unlearning effectiveness with its side effects (global stability and client performance change) in~\cref{Utilities and Strategies of Players}.
In~\cref{Two-Stage Decision Framework}, we propose a Stackelberg game to model the strategic interactions between the server and clients in the FU process.

\subsection{Utilities and Strategies of Players}\label{Utilities and Strategies of Players}

In our strategic FU framework, clients and the server interact through participation levels and payments. Specifically, each client $i \in \mathcal{N}$ determines their participation level $x_i \in [0,1]$, which represents the proportion of their local data (uniformly sampled from their local data distribution) used in the unlearning process. The dynamic client participation, characterized by the participation profile $\boldsymbol{x} = (x_1, \dots, x_N)$, affects FU outcomes.
To incentivize participation for FU, the server offers a payment scheme $\boldsymbol{p} = (p_1, \dots, p_N)$, where $p_i$ is the payment per unit of participation for client $i$.

\subsubsection{\textbf{Server's Decision Problem}}

The server seeks to achieve effective unlearning while mitigating the side effect (the stability deterioration), through strategic client participation in FU. 
To this end, it designs a payment scheme $\boldsymbol{p} = (p_1, \dots, p_N) \in \mathbb{R}_+^N$, where $p_i \in \mathbb{R}_+$ represents the price per unit of client $i$'s participation level. Given clients' participation strategies $\boldsymbol{x} = (x_1, \dots, x_N)$, the server aims to balance three objectives: minimizing the unlearning effectiveness metric $V$ to ensure target clients' removal, maintaining global stability $S$ to preserve system performance, and controlling the total payment costs. This trade-off is captured by maximizing server-side utility $U_\text{sr}$:
\begin{equation}\label{eq:server_utility}
\max_{\boldsymbol{p} \in \mathbb{R}_+^N} [U_\text{sr}(\boldsymbol{x}(\boldsymbol{p}), \boldsymbol{p}) \!:= \!-\lambda_v V(\boldsymbol{w}^{\boldsymbol{x}}) \!-\! \sum_{i=1}^N p_i x_i \!-\! \lambda_s S(\boldsymbol{w}^{\boldsymbol{x}})] \,, 
\end{equation}
where $\boldsymbol{w}^{\boldsymbol{x}}$ is the unlearned model resulting from client participation $\boldsymbol{x}$, $V(\cdot)$ and $S(\cdot)$ are the effectiveness and stability metrics defined in~\cref{def:fu_effective,def:fu_stab}, and $\lambda_v, \lambda_s > 0$ are weights balancing their relative importance. For notational simplicity, we use $V(\boldsymbol{x})$ and $S(\boldsymbol{x})$ to denote $V(\boldsymbol{w}^{\boldsymbol{x}})$ and $S(\boldsymbol{w}^{\boldsymbol{x}})$, with $\boldsymbol{x}(\boldsymbol{p}) = (x_1(p_1), \dots, x_N(p_N))$ representing the clients' responses to the payment scheme.

\subsubsection{\textbf{Client's Decision Problem}}

Each client $i \in \mathcal{N}$ determines their participation level $x_i \in [0,1]$ by considering (i) the \textit{payment received} from the server ($p_i x_i$), (ii) potential degradation in their local model performance $Q_i$ (defined in~\cref{def:fu_local}), which represents the \textit{local side effect} of FU, and (iii) the \textit{cost of participation}. 
Given the server's payment $p_i$ and other clients' strategies $\boldsymbol{x}_{-i}$, the client $i$ maximizes utility $U_i$:
\begin{equation}\label{eq:client_opt}
\max_{x_i \in [0,1]} \quad [U_i(x_i; \boldsymbol{x}_{-i}, p_i)  := p_i x_i - \lambda_q Q_i(\boldsymbol{w}^{\boldsymbol{x}}) - c_i x_i] \,, 
\end{equation} 
where $Q_i(\cdot)$ is the local performance change metric defined in~\cref{def:fu_local}, $\lambda_q > 0$ weights the importance of performance change, and $c_i > 0$ represents the cost per unit of participation. For notational simplicity, we use $Q_i(\boldsymbol{x})$ to denote $Q_i(\boldsymbol{w}^{\boldsymbol{x}})$. The solution to this problem determines the client's best response $x_i^*(p_i)$ to the server's payment scheme.

\subsection{Two-Stage Decision Framework}\label{Two-Stage Decision Framework}
We model this sequential decision-making between the server and remaining clients in FU as a two-stage Stackelberg game~\cite{bacsar1998dynamic}, where the server (leader) first determines the optimal payment scheme, and then the clients (followers) respond by choosing their optimal participation strategies. 

The Stackelberg equilibrium $(\boldsymbol{p}^*, \boldsymbol{x}^*)$ is characterized by:

\begin{definition}[Stackelberg Equilibrium]
A strategy profile $(\boldsymbol{p}^*, \boldsymbol{x}^*)$ constitutes a Stackelberg equilibrium if:

$\bullet$ The server's optimal payment scheme $\boldsymbol{p}^* = (p_1^*, \dots, p_N^*)$, given the best responses of the clients, maximizes its utility: $\boldsymbol{p}^* = \text{argmax}_{\boldsymbol{p}} U_\text{sr}(\boldsymbol{x}^*(\boldsymbol{p}), \boldsymbol{p})$.

$\bullet$ The clients' optimal strategies $\boldsymbol{x}^* \!=\!(x_1^*, \dots, x_N^*)$ maximize their individual utilities:
    $x_i^* \!=\! \text{argmax}_{x_i \in [0,1]} U_i(x_i, \boldsymbol{x}_{-i}^*; p_i^*)$, $\forall i \in \mathcal{N}$.\looseness=-1

\end{definition}

This Stackelberg equilibrium captures the strategic interdependence between the server's payment decisions and the clients' participation choices. 
The equilibrium payment scheme $\boldsymbol{p}^*$ and the resulting participation strategies $\boldsymbol{x}^*$ offer insights into the optimal design of incentives for the expected outcomes in terms of both individual and system-wide performance.

In the rest of the paper, 
we analyze this FU Stackelberg game, focusing on \textit{how data heterogeneity shapes the strategic interactions} between clients and the server. 
We develop a heterogeneity-aware framework that captures the impact of data heterogeneity on both client and server utilities in the FU process~\cref{Quantifying Heterogeneity Effects}. Then, we examine the clients' strategic behavior (\cref{sec:client_equilirium}) and server's payments for achieving desired system-wide outcomes (\cref{sec:serve_opt}).

\section{Quantifying Heterogeneity Effects}\label{Quantifying Heterogeneity Effects}

In FU, evaluating the impact of client participation presents unique challenges. Unlike FL, where participation primarily affects model convergence, in FU, the participation levels directly influence the unlearning effectiveness and the preservation of remaining knowledge. 
This creates two critical challenges specific to FU. First, the unlearned model's performance depends on the interplay between participating clients' data distributions and removed clients' distributions, where higher participation from similar clients may improve unlearning but risk model stability. 
Second, each client's local performance after unlearning depends on both their participation level and data heterogeneity relative to other participating clients and removed clients. 

To address these challenges, we need a systematic framework to analyze how data heterogeneity impacts FU from the server and client sides. 
In this section, we first introduce kernel mean embedding as a fundamental tool for measuring heterogeneity, then define heterogeneity at both client and system levels. 
We then establish theoretical bounds on how heterogeneity affects model performance, which forms the foundation for our heterogeneity-aware utility functions.

\subsection{Measuring Data Heterogeneity with Kernel Mean Embeddings}

To quantify data heterogeneity, we adopt kernel mean embedding (KME), which maps probability distributions into a reproducing kernel Hilbert space (RKHS). This approach enables efficient comparison of distributions through inner products in the RKHS, avoiding direct comparison of raw distributions.

\begin{definition}[Kernel Mean Embedding]
The kernel mean embedding of a distribution $\mathcal{D}$ is defined as:
\[
\mu = \mathbb{E}_{z \sim \mathcal{D}}[\phi(z)],
\]
where $\phi(z)$ is a feature map corresponding to the kernel function $k(\cdot, \cdot)$.
\end{definition}

We utilize KME to provide a mathematically tractable framework for quantifying data heterogeneity in FU, as it enables explicit bounds on model performance differences through RKHS norm calculations and facilitates theoretical analysis of strategic client behavior. 
While no single approach can fully capture all dimensions of real-world heterogeneity, 
KME captures the most statistically significant aspects of distribution differences 
that directly impact unlearning outcomes, allowing us to derive principled analysis.

The KME approach offers several crucial benefits for our mechanism design: (1) it provides \textit{a unified way to measure heterogeneity} across different data types and distributions without requiring direct access to client data, (2) it enables derivation of \textit{closed-form expressions} for client utilities and server objectives through kernel operations, allowing rigorous game-theoretic analysis, and (3) it offers principled ways to \textit{bound the impact of heterogeneity on unlearning performance} through kernel-based statistical learning theory. Unlike previous heuristic approaches, our KME framework can be extended with different kernel choices to capture various aspects of heterogeneity, providing a foundation that future work can build upon to address even more complex heterogeneity dimensions.\looseness=-1

\subsubsection{Client-Level Heterogeneity}
Let $\Omega = \{\mu_i\}_{i \in \mathcal{N}}$ denote the set of KMEs of clients' data distributions, where $\mu_i$ is the mean embedding for client $i$. For any pair of clients $i,j \in \mathcal{N}$, their data heterogeneity can be measured using the RKHS norm of the difference between their mean embeddings \(\|\mu_i - \mu_j\|_{\mathcal{H}}\). This client-level heterogeneity measure is crucial for analyzing how differences in data distributions affect both unlearning effectiveness and individual client performance. For instance, clients with high heterogeneity relative to removed clients may contribute more effectively to unlearning while better preserving their local performance.\looseness=-1

\subsubsection{System-Level Heterogeneity}~\label{system_level_h}
To analyze the collective impact of strategic client participation, we define a system-level heterogeneity with the aggregated KME:
\begin{equation}\label{eq:mixture_embedding}
\mu(\boldsymbol{x}) = \frac{\sum_{i \in \mathcal{N}} \alpha_i x_i \mu_i}{\sum_{i \in \mathcal{N}} \alpha_i x_i} \,,
\end{equation}
where $\alpha_i$ are client weights defined in~\cref{eq:fu_obj}, and $x_i \in [0,1]$ represents client $i$'s participation level. This system-level heterogeneity captures how the system's underlying data distribution evolves with dynamic participation, allowing us to analyze the impact of participation strategies on FU evaluations.

For the special case where all remaining clients fully participate ($x_i = 1, \forall i \in \mathcal{N}$), we denote \(
\mu_\mathcal{N} = \sum_{i \in \mathcal{N}} \alpha_i \mu_i
\).

\subsection{Heterogeneity-Aware Strategic Formulation}
To analyze how data heterogeneity affects unlearning performance, we need to quantify the relationship between distribution differences and model performance changes.  
We examine how distribution differences affect model performance through kernel mean embeddings. The following lemma establishes a fundamental relationship between distribution differences and model performance:

\begin{lemma}[Loss Difference Bound via Kernel Mean Embeddings]\label{lemma:loss_kernel_mean}
Let $\mathcal{H}$ be a reproducing kernel Hilbert space with kernel $k(\cdot,\cdot)$. Consider two probability distributions $\mathcal{D}_1$ and $\mathcal{D}_2$ with kernel mean embeddings $\mu_1 = \mathbb{E}_{z\sim \mathcal{D}_1}[k(\cdot,z)]$ and $\mu_2 = \mathbb{E}_{z \sim \mathcal{D}_2}[k(\cdot,z)]$. Let $w_1, w_2 \in \mathcal{H}$ be the minimizers of the regularized exponential family log-likelihood:
\[
w_i = \arg\min_{w \in \mathcal{H}} \left\{ -\mathbb{E}_{z\sim \mathcal{D}_i}[\log P_w(z)] + \frac{\lambda}{2}\|w\|^2_{\mathcal{H}} \right\}
\]
where $P_w(z) = \exp(\langle w, k(\cdot,z) \rangle_{\mathcal{H}} - A(w))$ and $A(w)$ is the log-partition function.
Then the following inequality holds:
\[
|\mathbb{E}_{z\sim \mathcal{D}_2}[\log P_{w_2}(z) - \log P_{w_1}(z)]| \leq C \|\mu_1 - \mu_2\|^2_{\mathcal{H}}
\]
where $\|\cdot\|_{\mathcal{H}}$ denotes the RKHS norm, and $C$ is a constant depending on the kernel and regularization parameter.
\end{lemma}

This lemma establishes a fundamental relationship between distribution differences (measured by KME distance) and model performance differences. In the context of FU, it enables us: 

$\bullet$ Predict how changes in client participation levels will affect model performance by measuring the resulting changes in the system-level heterogeneity  $\mu(\boldsymbol{x})$ mentioned in~\cref{system_level_h}, \textit{agnostic to the specific training procedure}.

$\bullet$ For individual clients, we can estimate performance changes by measuring the heterogeneity between their local distribution and the system-level heterogeneity. 

$\bullet$ Apply to a broad class of machine learning models, including commonly used algorithms like linear regression and support vector machines~(SVMs)~\cite{montesinos2022reproducing}, making it practically applicable for various FU scenarios. 

\begin{remark}[Heterogeneity-Based Bounds for FU Metrics]
In our FU framework, we use the negative log-likelihood as the evaluation (loss) function: $\ell(\boldsymbol{w}, z) = -\log P_{\boldsymbol{w}}(z)$. Under this choice, the difference in expected loss between two models $\boldsymbol{w}_1$ and $\boldsymbol{w}_2$ on a distribution $\mathcal{D}$ can be expressed as $\mathbb{E}_{z\sim \mathcal{D}}[\log P_{\boldsymbol{w}_2}(z) - \log P_{\boldsymbol{w}_1}(z)]$. Applying Lemma \ref{lemma:loss_kernel_mean}, we can bound our FU metrics in terms of distribution differences:

$\bullet$ Unlearning Effectiveness: The effectiveness metric $V(\boldsymbol{x})$ is bounded by the distributional difference between the ideal case (full participation) and the actual participation scenario:
\begin{equation}
V(\boldsymbol{x}) \leq C\|\mu_\mathcal{N} - \mu(\boldsymbol{x})\|^2_{\mathcal{H}}
\end{equation}
where $\mu_\mathcal{N}$ represents the distribution under full participation (corresponding to the retrained model $\boldsymbol{w}^r$), and $\mu(\boldsymbol{x})$ represents the distribution under participation strategy $\boldsymbol{x}$.

$\bullet$ Global Stability: The stability metric $S(\boldsymbol{x})$ is bounded by how much the participating clients' collective distribution deviates from the original training distribution:
\begin{equation}
S(\boldsymbol{x}) \leq C\|\mu_\mathcal{O} - \mu(\boldsymbol{x})\|^2_{\mathcal{H}}
\end{equation}
where $\mu_\mathcal{O}$ represents the distribution of all clients before unlearning (corresponding to the original model $\boldsymbol{w}^o$).

$\bullet$ Client Performance Change: For each client $i$, the performance change $Q_i(\boldsymbol{x})$ is bounded by:
\begin{equation}
Q_i(\boldsymbol{x}) \leq C\|\mu_i - \mu(\boldsymbol{x})\|^2_{\mathcal{H}} + \delta_i
\end{equation}
where $\mu_i$ represents client $i$'s local distribution, and $\delta_i = f_i(\boldsymbol{w}_i^*) - f_i(\boldsymbol{w}^o)$ accounts for the inherent gap between the client's local optimal model $\boldsymbol{w}_i^*$ and the global model $\boldsymbol{w}^o$.
\end{remark}

\noindent These theoretical bounds provide a robust framework for analyzing the impact of data heterogeneity on federated unlearning performance. By expressing the key performance metrics in terms of KME distances, we can now reformulate the strategic optimization problems faced by both the server and clients in a way that directly incorporates heterogeneity considerations.

\subsubsection{Heterogeneity-Aware Utility Functions.} 
Building on our framework that bounds performance metrics through distribution distances in RKHS, 
we can reformulate the optimization problems for both the server and clients to explicitly account 
for data heterogeneity effects.
The server's problem~\eqref{eq:server_utility}  becomes:

\begin{align}\label{opt:server}
\max_{\boldsymbol{p} \in \mathbb{R}_+^N} \left[ U_\text{sr}(\boldsymbol{x}(\boldsymbol{p}), \boldsymbol{p}) :=  -u_\text{server}(\boldsymbol{x}) - \sum_{i\in \mathcal{N}} p_i x_i \right] \,,
\end{align}
where $u_\text{server}(\boldsymbol{x}) = \hat{\lambda}_v \| \mu(\boldsymbol{x})-\mu_{\mathcal{N}}\|_{\mathcal{H}}^2 + \hat{\lambda}_s  \| \mu(\boldsymbol{x}) - \mu_\mathcal{O} \|_{\mathcal{H}}^2$, $\hat{\lambda}_v  = \lambda_v C$ and $\hat{\lambda}_s = \lambda_s C$, $\lambda_s, \lambda_v$ are weighting factors defined in~\cref{eq:server_utility}. 

Similarly, each client's decision problem~(\cref{eq:client_opt}) can be expressed as:
\begin{align}
\max_{x_i \in [0,1]} \left[U_i(x_i; \boldsymbol{x}_{-i}, p_i):= - u_i(\boldsymbol{x}) +  p_i x_i - c_i x_i\right]  \,,
\end{align}
where $u_i(\boldsymbol{x}) = \hat{\lambda}_q \| \mu(\boldsymbol{x})-\mu_{i}\|_{\mathcal{H}}^2$, $\hat{\lambda}_q = \lambda_q C$ with the weighting factor $\lambda_q$ defined in~\cref{eq:client_opt}. 

 These formulations directly capture how data heterogeneity affects decision-making: the server balances the unlearned model's distance from both the ideal retrained model ($\mu_{\mathcal{N}}$) and the original model ($\mu_\mathcal{O}$), while clients consider their individual data distribution's distance ($\mu_i$) from the aggregate distribution of participating clients ($\mu(\boldsymbol{x})$).

\subsubsection{Participation Strategies under Heterogeneity.} In the following lemmas and propositions, we explore the properties of \textit{performance impact function} $u_i(\boldsymbol{x})$, providing insights into how client heterogeneity and participation strategies influence performance outcomes in the FU process.
We begin by examining whether the performance change metric is non-increasing with respect to a client's participation level:

\begin{lemma}[Non-increasing Performance Impact]\label{lemma:monotonic_Q}
For any client $i \in \mathcal{N}$, given fixed strategies $\boldsymbol{x}_{-i}$ of other clients, the performance impact function $u_i(\boldsymbol{x})$ is strictly monotonically decreasing with respect to client~$i$'s participation level $x_i$, where $u_i(\boldsymbol{x}) = \hat{\lambda}_q \|\mu(\boldsymbol{x}) - \mu_i\|^2_{\mathcal{H}}$.
\end{lemma}

\begin{remark}
This lemma implies that client~$i$ benefits from increased participation $x_i$ in the FU process as it reduces $u_i$. 
Intuitively, participation helps mitigate the performance drop by aligning the FU system's data distribution with the client's.
Besides, if client~$i$'s data distribution is identical to that of other clients, $u_i$ remains constant regardless of participation level. In this scenario, the client's decision does not affect its local performance.
\end{remark}

Building on this lemma, we can quantify the change in performance impact when a client modifies their participation level:

\begin{corollary}[Participation Impact]\label{corollary:full_impact}
For any client $k \in \mathcal{N}$, given participation levels $x_k > x_k'$ and fixed strategies $\boldsymbol{x}_{-k}$, the change in performance impact is:
\begin{align*}
  \Delta u_k &= u_k(x_k, \boldsymbol{x}_{-k}) - u_k(x'_k, \boldsymbol{x}_{-k}) \\
  &= A\left(\frac{1}{(1 + \beta x_k)^2} - \frac{1}{(1 + \beta x'_k)^2}\right) \leq 0  \,,
\end{align*}
where $A = \|\hat{\mu} - \mu_k\|^2_{\mathcal{H}}$ and $\beta = a_k/(\sum_{j \neq k} a_j x_j)$
\end{corollary}

\begin{remark}\label{remark:large_datasets}
The benefit of participation (smaller $\Delta u_k$) increases as $\alpha_k$ increases, implying that clients with larger datasets have a potentially stronger incentive to participate. The participation of clients with a larger $\alpha_k$ (larger dataset) has a greater impact on shaping the unlearned model, potentially leading to a model that better reflects their local data distribution (\textit{e}.\textit{g}., smaller $u_k(x_k=1, \boldsymbol{x}_{-k} )$). However, larger datasets also introduce higher local computational costs $c_i$ to the utility.
\end{remark}

To characterize the performance change metric and its dependence on heterogeneity, we present the following~proposition:

\begin{proposition}[Performance Change Metric Properties]
\label{prop:fairness_metric_formal}

Consider two  $\Omega_1 = \{\mu_1, ..., \mu_k, ...\}$ and $\Omega_2 = \{\mu_1, ..., \mu'_k, ...\}$ differing only at index $k$. Consider participation strategies $x, x' \in [0,1]^n$ where $x_k > x'_k$ and $x_i = x'_i$ for all $i \neq k$. 

Then, if $\|\mu_k - \hat{\mu}\|_{\mathcal{H}} \geq \|\mu'_k - \hat{\mu}\|_{\mathcal{H}}$ where $\hat{\mu} = \sum_{i \neq k} \frac{a_i x_i}{\sum_{j \neq k} a_j x_j} \mu_i$:
$$u_i (x; \Omega_1) - u_i (x'; \Omega_1) \leq u_i (x; \Omega_2) - u_i (x'; \Omega_2)$$

\end{proposition}

\begin{remark}\label{remark:heterogeneity}
This proposition captures the relationship between a client's data heterogeneity and its incentive to participate in the FU process. 

Specifically, as client~$i$'s data distribution becomes more divergent from the FU system (represented by a larger $\|\mu_k - \hat{\mu}\|_{\mathcal{H}}$), the client $k$ gains more utility from joining the FU process (captured by a lower value of $u_i (x; \Omega_1) - u_i (x'; \Omega_1)$ under $x_k > x_k'$).
This indicates a higher incentive for client~$i$ to participate in the FU process. 
As suggested by \cref{lemma:monotonic_Q}, participation in the FU process helps mitigate potential performance drops by aligning the FU system's data distribution more closely with client~$i$. As a result, clients with more divergent data distributions have a stronger motivation to engage in the FU process.
\end{remark}

\section{Strategic Equilibrium in FU}
This section delves into the strategic interactions between clients and the server in FU. Our analysis unfolds in two key stages. 
In~\cref{sec:client_equilirium}, we analyze the clients' (followers) participation strategies that form a Nash equilibrium in response to a server payment scheme, proving the existence and uniqueness conditions and revealing how data heterogeneity influences client behaviors. 
In~\cref{sec:serve_opt}, we explore the server's (leader) optimal payment strategy, 
considering the anticipated Nash equilibrium responses of the clients. 
To efficiently solve this nonconvex optimization problem, 
we propose an efficient algorithm by transforming 
the server's decision problem into a quasiconvex optimization.

\subsection{Client Participation Strategies and Equilibrium}\label{sec:client_equilirium}
This section examines the strategic behavior of clients in the FU process, focusing on how data heterogeneity and incentive mechanisms influence client participation decisions. We analyze the Nash equilibria in the client-side game, providing insights into the stable outcomes of client participation strategies.

We begin by formally defining the Nash equilibrium concept in the context of our client-side game:

\begin{definition}[Nash Equilibrium among Clients]
Given the server's payment scheme $\boldsymbol{p} = (p_1, \dots, p_N)$, a strategy profile $\boldsymbol{x}^* = (x_1^*, \dots, x_N^*)$ is a Nash equilibrium among the clients if, for each client $i \in \mathcal{N}$:
\begin{equation}
U_i(x_i^*, \boldsymbol{x}_{-i}^*; p_i) \geq U_i(x_i, \boldsymbol{x}_{-i}^*; p_i), \text{ } \forall x_i \!\in\! [0, 1].
\end{equation}
\end{definition}
% The equilibrium states the stable outcomes of client participation strategies. 

To establish the existence of a Nash equilibrium, we first prove a key property of the client utility function:

\begin{lemma}[Concavity of Client Utility Function]
\label{lemma:concavity}
Given fixed strategies $\boldsymbol{x}_{-i}$ of other clients, the utility function $U_i$ of client~$i$ is concave in $x_i$ for $x_i \in [0, 1]$.
\end{lemma}

This concavity property is crucial as it ensures that each client's best response to others' strategies is well-defined. Building on this, we can now prove the existence of a Nash equilibrium in the client-side game:
\begin{proposition}[Existence of Nash Equilibrium]
\label{theorem:nash_existence}
Given the server's payment scheme $\boldsymbol{p} = (p_1, \dots, p_N)$, a Nash equilibrium exists in the client-side game. 
\end{proposition}

The existence of the Nash equilibrium ensures stable client strategies in response to the server's payment scheme. To fully understand the strategic landscape, we need to characterize the structure of these equilibria:

\begin{theorem}[Characterization of Nash Equilibrium]
\label{theorem:nash_characterization}
In any Nash Equilibrium $\boldsymbol{x}^*$, for each client $k \in \mathcal{N}$, given payment rate $p_k$, the equilibrium strategy $x_k^*$ is characterized as follows:
\begin{equation}\label{eq:x_opt}
x_k^*(p_k) = \begin{cases}
0 & \text{if } p_k < \underline{p_k}  \\
\left( \frac{\phi(\boldsymbol{x}^*_{-k})}{\alpha_k^2 (c_k-p_k)} \right)^{1/3}  - \frac{S_0}{\alpha_k} & \text{if } \underline{p_k} \leq p_k \leq \overline{p_k} \\
1 & \text{if } p_k > \overline{p_k}
\end{cases}
\end{equation}
where $\phi(\boldsymbol{x}^*_{-k}) = 2 \hat{\lambda}_q\| \sum_{i \neq k} \alpha_i x^*_i (\mu_i - \mu_k)\|_\mathcal{H}^2$, and the threshold payment are:
\begin{equation}\label{eq:threshold_p}
\underline{p_k} = c_k- \frac{\alpha_k \phi(\boldsymbol{x}_{-k}^*)}{S_0^{3}}, \quad \overline{p_k} = c_k - \frac{\alpha_k \phi(\boldsymbol{x}_{-k}^*)}{(S_0 + \alpha_k)^3} \,.
\end{equation}
\end{theorem}

This characterization reveals a threshold structure in the clients' equilibrium strategies, providing insights into understanding how clients will respond to different payment schemes and how their data heterogeneity influences their decisions.
In practical scenarios, it's desirable to have a unique equilibrium to ensure predictable outcomes. To this end, we provide a sufficient condition for the uniqueness of the Nash equilibrium:

\begin{theorem}[Sufficient Condition for Uniqueness of Nash Equilibrium]
\label{theorem:suff_uniqueness_nash}

For the Nash Equilibrium (NE) characterized in Theorem~\ref{theorem:nash_characterization}, if for  $\forall k \in \mathcal{N}$:
$
D < \frac{3 \alpha_k (1-\alpha_k)^2}{4}  \sqrt{ \frac{3 |c_k-p_k|}{\hat{\lambda}_q}} \,
$, 
where $D =\max_{i, j \in \mathcal{N}} \|\mu_i - \mu_j\|_{\mathcal{H}}$.
Then the NE is unique.\looseness=-1
\end{theorem}

\begin{remark}
The sufficient condition for the uniqueness of the Nash Equilibrium offers several game-theoretic insights:

1. Data Heterogeneity Bound: The uniqueness of the equilibrium is guaranteed when the maximum pairwise distance between client distributions (measured in $\mathcal{H}$-norm) is bounded above. In particular, as $\max_{i, j \in \mathcal{N}} \|\mu_i - \mu_j\|_{\mathcal{H}} \to 0$, the uniqueness condition is always satisfied.

2. Heterogeneity-Participation Trade-off: The bound on allowable heterogeneity is maximized when $\alpha_k \approx \frac{1}{3}$, which can be verified by maximizing the function $\alpha_k(1-\alpha_k)^2$. Conversely, systems dominated by a single client ($\alpha_i \rightarrow 1$) require near-homogeneous data distributions to guarantee uniqueness. 
For example, a system with highly skewed participation (\textit{e}.\textit{g}., 80\%-10\%-10\%) demands low heterogeneity for a unique equilibrium, while a more balanced distribution (\textit{e}.\textit{g}., 40\%-30\%-30\%) can tolerate higher levels of heterogeneity while still maintaining a unique equilibrium.

3. Incentive Mechanism for Heterogeneity: The absolute difference $|c_i-p_i|$ between participation cost and payment positively contributes to satisfying the condition.
In more heterogeneous environments, more aggressive pricing strategies (larger $|c_i-p_i|$) are necessary to ensure a unique~equilibrium.
\end{remark}

The above insights from the sufficient condition for uniqueness highlight the challenges posed by heterogeneity and the importance of carefully designed incentive mechanisms. 
For the special case of homogeneous clients, the strategic landscape simplifies considerably:
\begin{corollary}[Existence of Unique Nash Equilibrium for Homogeneous Clients]
In the FU client-side game with homogeneous remaining clients, a unique Nash equilibrium exists. 
The best response of client~$i$ is therefore:
\begin{equation}
BR_i(p_i) = \begin{cases}
1 & \text{if } p_i > c_i \\
0 & \text{if } p_i \leq c_i \\
\end{cases} \,.
\end{equation}
\end{corollary}
This corollary provides the special case of homogeneous clients in the FU client-side game, where clients do not experience performance reductions in FU, as all clients have similar data distributions. 
This best response function is independent of other clients' strategies, making it dominant.

\subsection{Server's Optimal Payment}\label{sec:serve_opt}

This section addresses the server's challenge of determining the optimal payment scheme for the server (leader) in the FU Stackelberg game. 
We first formulate the server's decision-making process as a nonconvex optimization problem. Then, we transform the initial nonconvex optimization problem into a quasiconvex optimization problem and introduce an efficient solution method.

The server's utility function involves a trade-off between minimizing $\|\mu(\boldsymbol{x})-\mu_{\mathcal{N}}\|_\mathcal{H}$ (to reduce V) and minimizing $\|\mu(\boldsymbol{x}) - \mu_{\mathcal{O}}\|_\mathcal{H}$ (to maintain S). These objectives can be conflicting, making the optimization problem non-trivial.

Given the clients' best responses characterized in~\cref{theorem:nash_characterization}, we reformulate~\cref{opt:server} as:
\begin{subequations}
\begin{align}
\textbf{P1:} \quad \min_{\boldsymbol{p}} \quad & u_\text{server}(\boldsymbol{x}^*(\boldsymbol{p})) \label{p1_us} \\
\text{s.t.} \quad & p_i \geq 0, \quad \forall i \in \mathcal{N} \label{p1:non_negative_p}\\
& \boldsymbol{p}^{\top} \boldsymbol{x}^*(\boldsymbol{p}) \leq B \label{p1:budget}
\end{align}
\end{subequations}
where $\boldsymbol{x}^*(\boldsymbol{p})$ is the clients' best responses to the payment scheme $\boldsymbol{p}$ as discussed in~\cref{theorem:nash_characterization}, and B is the budget.

The nonconvexity of \textbf{P1} presents a significant challenge for finding global optimal solutions efficiently. To address this, we transform \textbf{P1} into a \textit{quasiconvex} optimization problem, which allows for efficient solution methods.
Specifically, we establish the quasiconvexity of the server's utility function (\cref{theorem:quasiconvex_Us}) by linearizing $u_\text{server}$ with respect to $\mu(\boldsymbol{x})$ around $\mu_0 = \mu(\mathbf{x^*}(\mathbf{p_0}))$, where $\mathbf{x^*}(\mathbf{p_0})$ is the Nash equilibrium strategy profile for some initial payment scheme~$\mathbf{p_0}$:
\begin{equation}\label{eq:tilde_us}
\begin{aligned}
 & \tilde{u}_{server}(\mathbf{x^*}(\boldsymbol{p}))
 = u_\text{server}(\mathbf{x^*}(\mathbf{p_0}), \mathbf{p_0}) + \\
 & \quad 2 \langle \hat{\lambda}_v (\mu_0 - \mu_{\mathcal{N}}) + \hat{\lambda}_s (\mu_0- \mu_{\mathcal{O}}), \mu(\mathbf{x^*}(\boldsymbol{p})) - \mu_0 \rangle_{\mathcal{H}}
\end{aligned}
\end{equation}

\begin{lemma}[Quasiconvexity of $\tilde{u}_{server}$]
\label{theorem:quasiconvex_Us}
The function $\tilde{u}_{server}(\boldsymbol{x}^*(\boldsymbol{p}), \boldsymbol{p})$ is quasi-convex.
\end{lemma}

However, the budget constraint $\boldsymbol{p}^{\top} \boldsymbol{x}^*(\boldsymbol{p}) \leq B$ remains nonconvex due to the complex dependence of $\boldsymbol{x}^*(\boldsymbol{p})$ on $\boldsymbol{p}$. To address this challenge, we show that this function possesses a monotonicity property:

\begin{lemma}\label{lemma:monotonicity_px}
The function $\boldsymbol{p}^{\top} \boldsymbol{x}^*(\boldsymbol{p})$ is monotonically decreasing in $\boldsymbol{p}$ for $p_i \in [\underline{p_i}, \overline{p_i}]$, $\forall i \in \mathcal{N}$.
\end{lemma}

By Lemma \ref{lemma:monotonicity_px}, the budget constraint can be transformed into a set of linear constraints. For each client~$i$, there exists a unique $p_{B, i}$ such that $p_{B, i} x_i^*(p_{B, i}) = B_i$, where $B_i$ is client~$i$'s share of the total budget $B$. For $p_i < \underline{p_i}$, we have $x_i^* = 0$, so we can set $p_{B, i} = \underline{p_i}$. This allows us to transform the budget constraint into individual upper bounds on payments: $\boldsymbol{p}^{\top} \boldsymbol{x}^*(\boldsymbol{p}) \leq B \iff p_i \leq p_{B, i}$ for all $i$.
In our experiment, the total budget B is distributed proportionally based on clients' data weights ($B_i = B * \alpha_i / \sum_j \alpha_j$).

Therefore, we can reformulate \textbf{P1} as the following quasiconvex optimization problem:
\begin{subequations}
\begin{align}
\textbf{P2:} \quad \min_{\boldsymbol{p}} \quad & \tilde{u}_{server}(\mu(\mathbf{x^*}(\boldsymbol{p}))) \label{p3_us} \\
\text{s.t.} \quad & 0 \leq p_i \leq p_{B, i}, \quad \forall i \in \mathcal{N} \label{p2:bounds_p}
\end{align}
\end{subequations}
\textbf{P2} explicitly incorporates the heterogeneity constraint derived from our analysis, ensuring that the resulting payment scheme accounts for the critical role of data heterogeneity in FU outcomes. 
The transformation from the original nonconvex problem \textbf{P1} to a quasiconvex optimization \textbf{P2} introduces an approximation error that can be bounded as $O(\|\mu(\mathbf{x^*}(\boldsymbol{p})) - \mu_0\|^2_\mathcal{H})$. This error bound represents the squared distance between the current distribution and the linearization point in the RKHS, capturing the maximum possible deviation in the objective function. When $\|\mu(\mathbf{x^*}(\boldsymbol{p})) - \mu_0\|^2_\mathcal{H} \leq \varepsilon)$, the solution to the transformed problem is provably $\varepsilon$-optimal for the original problem.

\begin{algorithm}[t]
\caption{Heterogeneity-Aware Incremental Payment Optimization (HAIPO)}
\label{alg:haipo}
\small
\begin{algorithmic}[1]
\Require $N, \lambda_v, \lambda_s, \mu_{\mathcal{N}}, \mu_{\mathcal{O}}, B, \epsilon$
\Ensure Optimal payment scheme $\boldsymbol{p}^*$

% \State $H^* \gets \frac{\lambda_v C H_{\mathcal{N}} + \lambda_s C H_{\mathcal{O}}}{\lambda_v C + \lambda_s C}$
% \State Initialize $\boldsymbol{p}_B$. 
\State // Compute Budget Upper Bounds $\boldsymbol{p}_B$\label{line:budget_up_start}
\For{$i = 1$ to $N$}
    \State $p_{B,i} \gets \textsc{BisectionSearch}(p_i x_i^*(p_i) = B_i)$\label{line:bisection_search}
\EndFor

\State $\boldsymbol{p}_\text{low} \gets \mathbf{0}, \boldsymbol{p}_\text{high} \gets \boldsymbol{p}_B$\label{line:budget_up_end}
\State $k \gets 0$, $\boldsymbol{p}_0 \gets \frac{1}{2}(\boldsymbol{p}_\text{low} + \boldsymbol{p}_\text{high})$ \label{line:main_start}
\Repeat
    \State $\boldsymbol{x}^*_k \!\gets\!\textsc{NashEquilibrium}(\boldsymbol{p}_k)$~Computed by~\cref{theorem:nash_characterization}
    \State $\mu_k \gets \mu(\boldsymbol{x}^*_k)$
    \State $\tilde{u}_{server}(\boldsymbol{x}^*_k(\boldsymbol{p}_k)) \gets$ Equation~\eqref{eq:tilde_us}
    % \State // Solve the quasiconvex optimization problem
    \State $\boldsymbol{p}_{k+1} \gets \textsc{QuasiconvexOptimization}(\tilde{u}_{server}, \boldsymbol{p}_\text{low}, \boldsymbol{p}_\text{high})$\label{line:quasiconvexopt}
    \State \textbf{if  }{$\|\boldsymbol{p}_{k+1} - \boldsymbol{p}_k\| < \epsilon$}: \textbf{break}
    \State $k \gets k + 1$
\Until{$k > \text{MaxIterations}$}
\State \Return $\boldsymbol{p}^*\gets \boldsymbol{p}_k$ \label{line:main_end}
\end{algorithmic}
\end{algorithm}

To solve \textbf{P2}, we propose Heterogeneity-Aware Incremental Payment Optimization (HAIPO), detailed in Algorithm \ref{alg:haipo}. 
Specifically, we employ a two-stage approach:

1. \textit{Budget Upper Bound Computation}: We first determine the budget upper bounds $\boldsymbol{p}_B$ for each client using bisection search~(Lines \ref{line:budget_up_start}-\ref{line:budget_up_end}). Specifically, \textsc{BisectionSearch}~(Line~\ref{line:bisection_search}) find the unique threshold payment $p_{B,i}$ that satisfies $p_{B,i}x_i^*(p_{B,i}) = B_i$ by iteratively narrowing the search interval $[0, B_i]$, where $x_i^*(\cdot)$ is characterized in~\cref{theorem:nash_characterization} and the uniqueness of $p_{B,i}$ is guaranteed by~\cref{lemma:monotonicity_px}.

2. \textit{Optimal Payment Search}: We then utilize bisection search to efficiently solve the quasiconvex optimization problem (Lines \ref{line:main_start}-\ref{line:main_end}). Specifically, \textsc{QuasiconvexOptimization} (Line~\ref{line:quasiconvexopt}) solves a sequence of convex feasibility problems by testing level sets $\{p: \tilde{u}_{server}(x^*(p)) \leq \alpha\}$ to find the optimal payment scheme $p^*$ that minimizes the transformed objective $\tilde{u}_{server}$ while satisfying the budget constraints $p_i \leq p_{B,i}$.
The complexity is $O(\log(\frac{1}{\epsilon}))$, where $\epsilon$ is the desired precision.

\vspace{-0.5em}
\section{Experiments}
In this section, we empirically evaluate our heterogeneity-aware incentive mechanism for FU. 
In the following, we first describe our experimental setup, 
including datasets, models, and evaluation metrics. 
Then, we provide a detailed analysis of the results.
\vspace{-0.5em}
\subsection{Experimental Setup}

\subsubsection{Datasets and Models} We conducted experiments using the BERT model~\cite{devlin2018bert} for \textit{text classification} on the AG\_NEWS dataset~\cite{zhang2015character}, and the ResNet model~\cite{he2016deep} for \textit{image classification} on the CIFAR-100 dataset~\cite{Krizhevsky09learningmultiple}.

\subsubsection{Data Heterogeneity and Parameters}
To simulate varying degrees of data heterogeneity, we utilize a Dirichlet distribution to allocate both \textit{label distribution} and \textit{data quantities} across clients. We employ three different Dirichlet concentration parameters ($\beta$): 0.2, 0.5, and 0.8, where the lower $\beta$ values result in more heterogeneous data distributions among clients.
 For each client $i$, the empirical KME $\hat{\mu}_i = \frac{1}{n_i} \sum_{j=1}^{n_i} k(\cdot, x_j^i)$ using the RBF kernel $k(x,y) = \exp(-\|x-y\|^2/2\sigma^2)$, where the bandwidth $\sigma$ is determined using the median heuristic $\sigma^2 = \text{median}\{\|x_i - x_j\|^2\}$ across all pairwise distances in the dataset. 

Additionally,
following \cite{ding2023incentive}, we consider client~$i$'s participation cost as $c_i = \gamma n_i$, where $n_i$ is local data size and $\gamma$ set as $10/n_\mathcal{O}$ is a scaling factor, with $n_\mathcal{O}$ defined in~\cref{eq:fl_obj}. 

For weighting parameters in the utility functions, we set $\lambda_v = \lambda_s = \lambda_q = 1$ as default to give equal importance to unlearning effectiveness, global stability, and local performance changes.\looseness=-1

\subsubsection{Evaluation Metrics} We evaluated our method using three key metrics (\textbf{in accuracy}):
\begin{itemize}
    \item \textit{Global Stability ($\tilde{S}$)}: Measures the overall model performance for original FL system. \textit{Higher} values indicate better retention of general model performance after FU.
    \item \textit{Unlearning Effectiveness ($\tilde{V}$)}: Assesses the unlearning effectiveness on remaining clients. \textit{Higher} values suggest better FU effectiveness.
    \item \textit{Client Performance Change ($\tilde{Q}$)}: Captures the maximum accuracy degradation among remaining clients (\textit{e}.\textit{g}., -11\% indicates the worst-affected client's performance drop).
\end{itemize}

\subsubsection{Baselines} To evaluate the performance of our proposed method, we compared it against the following baselines:

\begin{itemize}
    \item \textit{Retrain}: The model is fully retrained from scratch using only the remaining clients' data.
    \item \textit{Uniform}: A uniform price scheme where all clients receive the same payment for participation.
    \item \textit{FU Methods}: Three FU approaches including FedEraser~\cite{liu2021federaser}, MoDe~\cite{zhao2023federated}, and QuickDrop~\cite{dhasade2023QuickDrop}.
\end{itemize}

\begin{table}[t]
    \centering
    \scriptsize
    \setlength{\tabcolsep}{4pt}
    \caption{Performance Comparison compared to Retrain baseline. Our method consistently outperforms baselines by balancing stability, unlearning effectiveness, and client performance. \small{
    (\textit{Note}: Retrain values show absolute performance, while all other methods show relative differences from retrain.})}
    \label{tab:results}
    \begin{adjustbox}{width=\columnwidth}
    \begin{tabular}{c|c|c|c|c|c}
    \hline
    Model-Dataset & $\beta$ & Method & \multicolumn{3}{c}{Performance Metrics} \\
    \cline{4-6}
     &  &  & $\tilde{V}$ & $\tilde{S}$ & $\tilde{Q}$ \\
    \hline
    \multirow{15}{*}{\begin{tabular}{c}Bert-\\AG\_NEWS\end{tabular}} &  \multirow{5}{*}{0.2} & \textit{retrain (baseline)} & 92.5\% & 95.0\% & -7.54\% \\
    \cline{3-6}
    & & uniform & +1.65\% & -2.05\% & +5.54\% \\
    & & FedEraser & 0.00\% & 0.00\% & 0.00\% \\
    & & MoDe & +2.40\% & -7.22\% & +2.22\% \\
    & & QuickDrop & -3.51\% & -13.26\% & -2.02\% \\
    & & \textbf{Ours} & \underline{-1.92\%} & \textbf{-1.37\%} & \textbf{+5.98\%} \\
    \cline{2-6}
    &  \multirow{5}{*}{0.5} & \textit{retrain (baseline)} & 87.7\% & 89.6\% & -11.05\% \\
    \cline{3-6}
    & & uniform & +5.45\% & +4.42\% & +9.05\% \\
    & & FedEraser & 0.00\% & 0.00\% & 0.00\% \\
    & & MoDe & +3.06\% & -6.61\% & +2.12\% \\
    & & QuickDrop & -3.03\% & -13.15\% & -2.29\% \\
    & & \textbf{Ours} & \underline{+5.47\%} & \textbf{+4.10\%} & \textbf{+10.05\%} \\
    \cline{2-6}
    &  \multirow{5}{*}{0.8} & \textit{retrain (baseline)} & 93.5\% & 91.5\% & -8.49\% \\
    \cline{3-6}
    & & uniform & +0.95\% & +2.81\% & +7.38\% \\
    & & FedEraser & 0.00\% & 0.00\% & 0.00\% \\
    & & MoDe & +1.93\% & -6.98\% & +2.20\% \\
    & & QuickDrop & -2.99\% & -13.52\% & -2.00\% \\
    & & \textbf{Ours} & \underline{+1.14\%} & \textbf{+2.84\%} & \textbf{+7.68\%} \\
    \hline
    \multirow{15}{*}{\begin{tabular}{c}Resnet-\\CIFAR100\end{tabular}} &  \multirow{5}{*}{0.2} & \textit{retrain (baseline)} & 50.37\% & 52.02\% & -9.62\% \\
    \cline{3-6}
    & & uniform & +4.00\% & +1.15\% & +5.56\% \\
    & & FedEraser & 0.00\% & 0.00\% & 0.00\% \\
    & & MoDe & +2.24\% & -5.49\% & +2.07\% \\
    & & QuickDrop & -2.87\% & -13.17\% & -2.44\% \\
    & & \textbf{Ours} & \underline{+8.82\%} & \textbf{+6.23\%} & \textbf{+6.62\%} \\
    \cline{2-6}
    &  \multirow{5}{*}{0.5} & \textit{retrain (baseline)} & 54.41\% & 54.36\% & -8.00\% \\
    \cline{3-6}
    & & uniform & +0.85\% & +0.86\% & +4.00\% \\
    & & FedEraser & 0.00\% & 0.00\% & 0.00\% \\
    & & MoDe & +1.32\% & -6.45\% & +2.19\% \\
    & & QuickDrop & -2.74\% & -11.27\% & -2.26\% \\
    & & \textbf{Ours} & \underline{+1.66\%} & \textbf{+1.67\%} & \textbf{+5.75\%} \\
    \cline{2-6}
    &  \multirow{5}{*}{0.8} & \textit{retrain (baseline)} & 54.89\% & 56.65\% & -5.00\% \\
    \cline{3-6}
    & & uniform & -1.44\% & +0.38\% & +3.00\% \\
    & & FedEraser & 0.00\% & 0.00\% & 0.00\% \\
    & & MoDe & +2.51\% & -6.76\% & +2.06\% \\
    & & QuickDrop & -2.41\% & -12.18\% & -2.21\% \\
    & & \textbf{Ours} & \underline{+2.85\%} & \textbf{+0.89\%} & \textbf{+4.00\%} \\
    \hline
    \end{tabular}
    \end{adjustbox}
    
    \end{table}
    
\subsection{Experimental Results} 
First, we present a comparative analysis of our proposed method's overall performance against the baselines. Then, we provide a detailed examination of client participation dynamics to gain deeper insights into the effectiveness of our heterogeneity-aware incentive mechanism.
In our experiments, clients indexed [0, 1, 2] are the removed clients.

\subsubsection{\textbf{Overall Performance Comparison}}

We conduct experiments across various heterogeneity levels and dataset-model combinations, presenting the results in~\cref{tab:results}.

\noindent $\bullet$ \textbf{Strategic Performance Trade-offs}: Our method demonstrates balanced performance across all metrics, while competitors typically excel in at most one dimension. With Bert-AG\_NEWS ($\beta=0.8$), we achieve a well-balanced performance: +2.84\% in stability and +7.68\% in client performance. In contrast, FedEraser achieves perfect unlearning but fails to mitigate side effects.

\noindent $\bullet$ \textbf{Mitigation of unlearning's side effects}: 
Our approach consistently delivers the strongest client performance improvements ($\tilde{Q}$), 
exceeding all baselines in every scenario tested: +10.05\% for Bert-AG\_NEWS ($\beta=0.5$) and +6.62\% 
for Resnet-CIFAR100 ($\beta=0.2$). This demonstrates our incentive mechanism's effectiveness in minimizing negative impacts on remaining clients while maintaining appropriate unlearning performance.

\noindent $\bullet$ \textbf{Robustness to heterogeneity}: 
Notably, in both high and low heterogeneity scenarios ($\beta=0.2\text{ and }0.8$), for Bert-AG\_NEWS, 
our method maintains positive gains across side effects.
This demonstrates our approach's robustness in 
different heterogeneous environments.

\noindent $\bullet$ \textbf{Computational Trade-offs}:
\revision{Our method achieves substantial computational efficiency compared to complete retraining, 
as shown by the average runtime comparison in Table~\ref{tab:computation} across various settings. 
In contrast to existing FU baselines that prioritize minimal overhead, 
our approach introduces a principled optimization procedure explicitly designed to mitigate the inherent side effects in FU.
As shown Table~\ref{tab:results}, our method achieves better global stability ($\tilde{S}$) and client performance ($\tilde{Q}$). 
By integrating system-level robustness, our method offers a more practical approach for real-world FU scenarios for regulated industries like healthcare and finance, where both compliance with removal requests and system reliability are essential.}

These results demonstrate that our heterogeneity-aware incentive mechanism effectively 
navigates heterogeneity-induced trade-offs in FU.
This balanced approach represents 
an advancement for practical FU mechanisms beyond the right-to-be-forgotten.

\begin{table}[t]
\centering
\small
\caption{Computational Cost Comparison (in minutes).}
\label{tab:computation}
\begin{tabular}{l|c|c|c|c|c}
\hline
Dataset & Retrain & FedEraser & MoDe & QuickDrop & Ours \\
\hline
CIFAR-100 & 11.5 & 4.6 & 3.5 & 2.3 & 8.2 \\
AG\_NEWS & 157.2 & 82.9 & 47.2 & 31.4 & 96.5 \\
\hline
\end{tabular}
\end{table}

\subsubsection{\textbf{Client Participation Dynamics}}

\begin{figure}[t]
    \centering    \includegraphics[width=\linewidth]{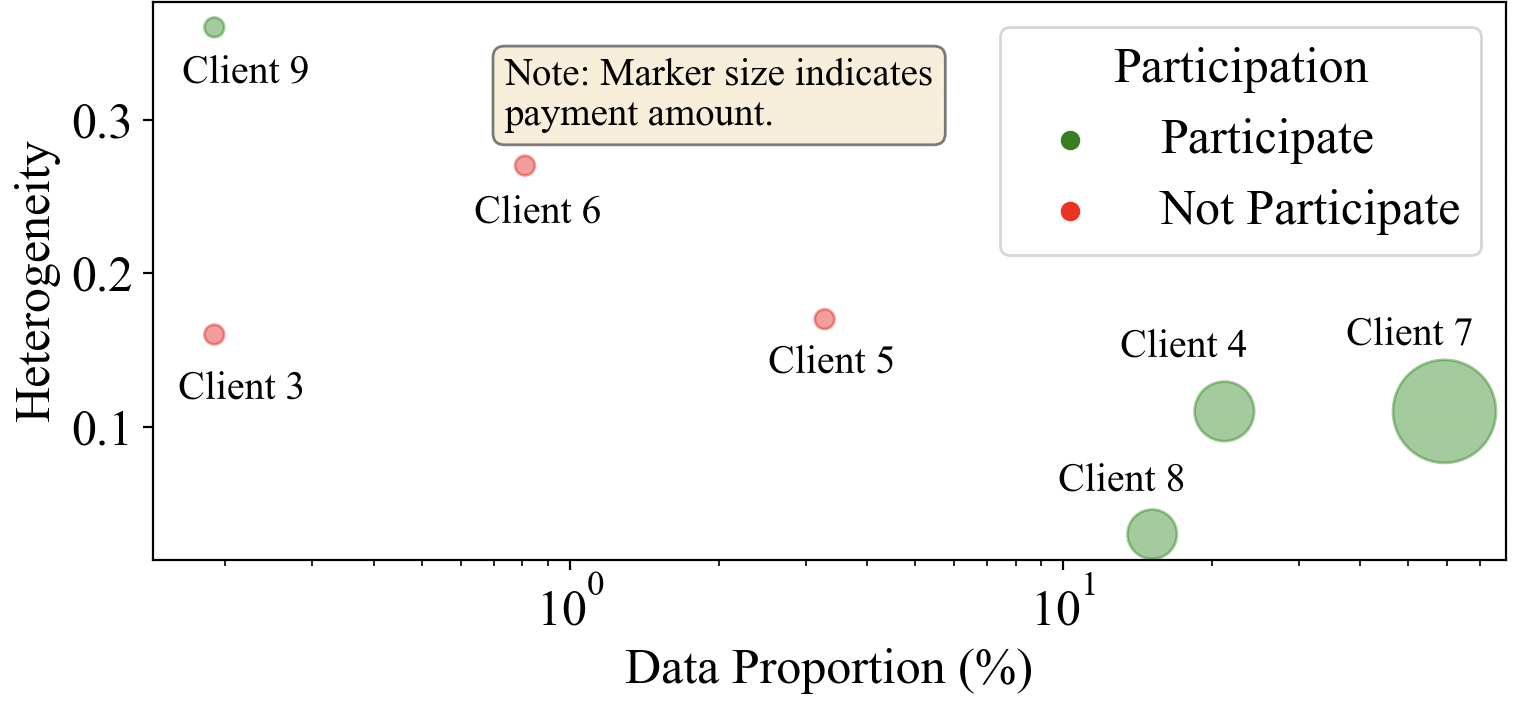}
    \caption{Client characteristics and participation: 
    clients with large data volume and high heterogeneity are incentivized to join FU. 
    (\small{\textit{Note}: Marker size indicates the server's payment amount for each client.})
    }
    \Description{Client characteristics and participation: clients with large data volume and high heterogeneity are incentivized to join FU. Marker size indicates the server's payment amount for each client.}
\label{fig:dynamic_participation}
\end{figure}

% \begin{table}[t]
% \caption{Client Characteristics and Participation}
% \label{tab:clientdata}
% \centering
% \begin{tabular}{c|r|r|r|c}
% \hline
% $i$ & Weight (\%) & Heterogeneity $H_i$ & Payment $p_i$ & Participation $x_i$ \\
% \hline
% 3 & 0.19 & 0.16 & 0 & 0 \\
% 4 & 21.19 & 0.11 & 2.05 & 1 \\
% 5 & 3.28 & 0.17 & 0 & 0 \\
% 6 & 0.81 & 0.27 & 0 & 0 \\
% 7 & 59.20 & 0.11 & 6.62 & 1 \\
% 8 & 15.13 & 0.03 & 1.33 & 1 \\
% 9 & 0.19 & 0.36 & 0 & 1 \\
% \hline
% \end{tabular}
% \end{table}
To better understand the dynamics of client participation in the unlearning process, we analyze the individual client characteristics and their corresponding decisions in the Bert-AG\_NEWS experiment with $\beta  = 0.2$ (highly heterogeneous setting), detailed in~\cref{fig:dynamic_participation}. In this Figure, \textit{data proportio}n represents a client's data quantity relative to all remaining clients, and \textit{heterogeneity} $\| \mu_i - \mu_\mathcal{R}\|_{\mathcal{H}}$ is measured between client~$i$'s data distribution and that of the removed clients $\mu_\mathcal{R} = \sum_{j \in \mathcal{R}} \alpha_j \mu_j / \sum_{j \in \mathcal{R}} \alpha_j$.

\noindent $\bullet$ \textbf{Data Volume Impact}: Clients 3, 5, and 6, with relatively small data volumes (0.19\%, 3.28\%, and 0.81\%, respectively), do not participate in the unlearning process. This aligns with the insight from \cref{remark:large_datasets}, which suggests that clients with larger datasets have a stronger incentive to participate due to their potential for a more significant impact on the model.

\noindent $\bullet$ \textbf{Heterogeneity Influence}: Despite having the same small data volume as Client 3 (0.19\%), Client 9 chooses to participate. This can be attributed to its significantly higher heterogeneity (0.36 compared to 0.16 for Client 3). This observation aligns with \cref{remark:heterogeneity}, which indicates that clients with data distributions more divergent from the FU system gain more utility from joining the FU process.

\noindent $\bullet$ \textbf{Payment Allocation}: The server allocates higher payments to clients with larger data volumes (Clients 4 and 7), recognizing their potential higher cost to join the unlearning process. However, the payment is not strictly proportional to the data volume, suggesting that other factors (such as heterogeneity) also play a role in determining the optimal payment scheme.

The selective participation strategy not only helps to achieve effective unlearning but also potentially reduces communication and computation costs by involving only the most valuable clients in the process. 
Only four of the seven remaining clients participate in the unlearning process ($x_i = 1$), demonstrating
the effectiveness of our approach in incentivizing participation from clients.

\section{Conclusion}
In this paper, we present a novel framework for incentive design in FU that addresses the complex interplay between system-wide objectives of unlearning effectiveness and global stability with individual client interests.
% Our approach, based on a Two-Stage Decision Framework, is the first to comprehensively model the strategic interactions between a server and participating clients 
% We introduced a heterogeneity-aware performance evaluation that captures the nuanced impacts of data heterogeneity on unlearning outcomes and client decisions. 
Through rigorous theoretical analysis, we established the existence and uniqueness conditions for the Nash equilibrium among clients, providing valuable insights for designing stable incentive mechanisms in FU. 
Additionally, we developed an efficient solution for the server. 
Our extensive experimental results validate our approach across varying settings, demonstrating significant improvements in both unlearning performance metrics and computational efficiency, with runtime reductions of up to 38.6\% compared to complete retraining.

\bibliographystyle{ACM-Reference-Format}
\bibliography{ref}

\appendix
\onecolumn

\section{Proof for~\cref{lemma:loss_kernel_mean}} \label{app:lemma:loss_kernel_mean}
\textbf{Lemma \ref{lemma:loss_kernel_mean}} (Loss Difference Bound via Kernel Mean Embeddings).
Let $\mathcal{H}$ be a reproducing kernel Hilbert space with kernel $k(\cdot,\cdot)$. Consider two probability distributions $\mathcal{D}_1$ and $\mathcal{D}_2$ with kernel mean embeddings $\mu_1 = \mathbb{E}_{z\sim \mathcal{D}_1}[k(\cdot,z)]$ and $\mu_2 = \mathbb{E}_{z \sim \mathcal{D}_2}[k(\cdot,z)]$. Let $w_1, w_2 \in \mathcal{H}$ be the minimizers of the regularized exponential family log-likelihood:
\[
w_i = \arg\min_{w \in \mathcal{H}} \left\{ -\mathbb{E}_{z\sim \mathcal{D}_i}[\log P_w(z)] + \frac{\lambda}{2}\|w\|^2_{\mathcal{H}} \right\}
\]
where $P_w(z) = \exp(\langle w, k(\cdot,z) \rangle_{\mathcal{H}} - A(w))$ and $A(w)$ is the log-partition function.
Then the following inequality holds:
\[
|\mathbb{E}_{z\sim \mathcal{D}_2}[\log P_{w_2}(z) - \log P_{w_1}(z)]| \leq C \|\mu_1 - \mu_2\|^2_{\mathcal{H}}
\]
where $\|\cdot\|_{\mathcal{H}}$ denotes the RKHS norm, and $C$ is a constant depending on the kernel and regularization parameter.

\begin{proof}
    1) First, we express the log-likelihood difference:
\begin{align*}
& \mathbb{E}_{d\sim \mathcal{D}_2}[\log P_{w_2}(z) - \log P_{w_1}(z)] \\
&= \mathbb{E}_{d\sim \mathcal{D}_2}[\langle w_2, k(\cdot,z) \rangle_{\mathcal{H}} - A(w_2) - \langle w_1, k(\cdot,z) \rangle_{\mathcal{H}} + A(w_1)] \\
&= \langle w_2, \mu_2 \rangle_{\mathcal{H}} - \langle w_1, \mu_2 \rangle_{\mathcal{H}} - [A(w_2) - A(w_1)]
\end{align*}

The second equation is obtained by the linearity of expectation and the definition of kernel mean embedding.

2) For the regularized log-likelihood optimization:
\[
\min_{w \in \mathcal{H}} \left\{ -\mathbb{E}_{d\sim \mathcal{D}_i}[\log P_w(z)] + \frac{\lambda}{2}\|w\|^2_{\mathcal{H}} \right\}
\]
The first-order optimality condition gives:
\(
-\mu_i + \nabla A(w_i) + \lambda w_i = 0
\).
This implies for $i \in \{1,2\}$:
\(
\nabla A(w_i) = \mu_i - \lambda w_i
\).

3) Since the log-partition function $A(w)$ is $\kappa$-convex ($\kappa > 0$):
\(
\langle \nabla A(w_2) - \nabla A(w_1), w_2 - w_1 \rangle_{\mathcal{H}} \geq \kappa\|w_2 - w_1\|^2_{\mathcal{H}}
\).

Substituting the expressions from step 2:
\begin{align*}
% \langle \mu_2 - \lambda w_2 - (\mu_1 - \lambda w_1), w_2 - w_1 \rangle_{\mathcal{H}} &\geq \kappa\|w_2 - w_1\|^2_{\mathcal{H}} \\
\langle \mu_2 - \mu_1, w_2 - w_1 \rangle_{\mathcal{H}} - \lambda\|w_2 - w_1\|^2_{\mathcal{H}} &\geq \kappa\|w_2 - w_1\|^2_{\mathcal{H}}
\end{align*}

By Cauchy-Schwarz inequality:
\(
\|\mu_2 - \mu_1\|_{\mathcal{H}} \cdot \|w_2 - w_1\|_{\mathcal{H}} \geq (\lambda + \kappa)\|w_2 - w_1\|^2_{\mathcal{H}}
\).
Therefore:
\(
\|w_2 - w_1\|_{\mathcal{H}} \leq \frac{1}{\lambda + \kappa}\|\mu_2 - \mu_1\|_{\mathcal{H}}
\).

4) Finally, applying Cauchy-Schwarz to the likelihood difference:
\begin{align*}
& |\mathbb{E}_{d\sim \mathcal{D}_2}[\log P_{w_2}(z) - \log P_{w_1}(z)]| \\ 
&\leq \|w_2 - w_1\|_{\mathcal{H}} \cdot \|\mu_2 - \mu_1\|_{\mathcal{H}} \leq \frac{1}{\lambda + \kappa}\|\mu_2 - \mu_1\|^2_{\mathcal{H}}
\end{align*}

Thus, we have shown that:
\(
|\mathbb{E}_{d\sim \mathcal{D}_2}[\log P_{w_2}(z) - \log P_{w_1}(z)]| \leq C\|\mu_2 - \mu_1\|^2_{\mathcal{H}}
\),
where $C = \frac{1}{\lambda + \kappa}$.
\end{proof}

\section{Proof for \cref{lemma:monotonic_Q}}\label{proof:lemma:monotonic_Q}
\textbf{Lemma \ref{lemma:monotonic_Q}} (Non-increasing Performance Impact).
For any client $i \in \mathcal{N}$, given fixed strategies $\boldsymbol{x}_{-i}$ of other clients, the performance impact function $u_i(\boldsymbol{x})$ is strictly monotonically decreasing with respect to client~$i$'s participation level $x_i$, where $u_i(\boldsymbol{x}) = \hat{\lambda}_q \|\mu(\boldsymbol{x}) - \mu_i\|^2_{\mathcal{H}}$.

\begin{proof}
Firstly, consider a set of distributions $\{\mathcal{D}_i\}$ with corresponding RKHS embeddings $\{\mu_i\}$. Let \(x_i \in [0,1]\) and weights \(a_i\) satisfy \(\sum_i a_i = 1\).  Let \(y_i = \frac{a_i x_i}{\sum_j a_j x_j}\), and \(S = \sum_j a_j x_j\). The weighted mean kernel embedding mentioned in~\cref{eq:mixture_embedding} becomes:
\(\mu(x) = \sum_i y_i \mu_i  = \sum_i \frac{a_i x_i}{S} \mu_i\).

1. Express the squared RKHS distance:
   \[\|\mu(x) - \mu_k\|^2_{\mathcal{H}} = \langle\mu(x), \mu(x)\rangle_{\mathcal{H}} - 2\langle\mu(x), \mu_k\rangle_{\mathcal{H}} + \langle\mu_k, \mu_k\rangle_{\mathcal{H}}\]

2. Compute derivatives of individual terms:
   \[\frac{\partial}{\partial x_k}\langle\mu(x), \mu(x)\rangle_{\mathcal{H}} = \frac{2\alpha_k}{S}[-\langle\mu(x), \mu(x)\rangle_{\mathcal{H}} + \langle\mu(x), \mu_k\rangle_{\mathcal{H}}]\]
   \[\frac{\partial}{\partial x_k}(-2\langle\mu(x), \mu_k\rangle_{\mathcal{H}}) = \frac{2\alpha_k}{S}[\langle\mu(x), \mu_k\rangle_{\mathcal{H}} - \langle\mu_k, \mu_k\rangle_{\mathcal{H}}]\]

3. The partial derivative of \(\|\mu(x) - \mu_k\|^2_{\mathcal{H}}\) with respect to \(x_k\) is:
\(\frac{\partial}{\partial x_k} \|\mu(x) - \mu_k\|^2_{\mathcal{H}} = -\frac{2\alpha_k}{S}\|\mu(x) - \mu_k\|^2_{\mathcal{H}}\).
Since \(\alpha_k > 0\) and \(S > 0\), the derivative $\frac{\partial}{\partial x_k} \|\mu(x) - \mu_k\|^2_{\mathcal{H}} \leq 0$.
\end{proof}

\section{Proof for~\cref{prop:fairness_metric_formal}}\label{app:proof_prop:fairness_metric_formal}

\textbf{Proposition \ref{prop:fairness_metric_formal}} (Performance Change Metric Properties).
Consider two  $\Omega_1 = \{\mu_1, ..., \mu_k, ...\}$ and $\Omega_2 = \{\mu_1, ..., \mu'_k, ...\}$ differing only at index $k$. Consider participation strategies $x, x' \in [0,1]^n$ where $x_k > x'_k$ and $x_i = x'_i$ for all $i \neq k$. 

Then, if $\|\mu_k - \hat{\mu}\|_{\mathcal{H}} \geq \|\mu'_k - \hat{\mu}\|_{\mathcal{H}}$ where $\hat{\mu} = \sum_{i \neq k} \frac{a_i x_i}{\sum_{j \neq k} a_j x_j} \mu_i$:
$$u_i (x; \Omega_1) - u_i (x'; \Omega_1) \leq u_i (x; \Omega_2) - u_i (x'; \Omega_2)$$

\begin{proof}
    
For $u_i (\boldsymbol{x}; \Omega_1) - u_i (\boldsymbol{x}'; \Omega_1) \leq u_i (\boldsymbol{x}; \Omega_2) - u_i (\boldsymbol{x}'; \Omega_2)$, it is equivalent to $\|\mu(\boldsymbol{x}; \Omega_1) - \mu_k\|_{\mathcal{H}}^2 - \|\mu(\boldsymbol{x}'; \Omega_1) - \mu_k\|_{\mathcal{H}}^2 \leq \|\mu(\boldsymbol{x}; \Omega_2) - \mu'_k\|_{\mathcal{H}}^2 - \|\mu(\boldsymbol{x}'; \Omega_2) - \mu'_k\|_{\mathcal{H}}^2$. 
Let $S = \sum_{j \neq k} a_j x_j$ and $\beta = \frac{a_k}{S}$. Then:
For any $\mu \in \{\mu_k, \mu'_k\}$, the aggregated model can be written as:
   $\mu(\boldsymbol{x}; \Omega) = \frac{\hat{\mu} + \beta x_k\mu}{1 + \beta x_k}$.

Consider the distance term:
   $\|\frac{\hat{\mu} + \beta x_k\mu}{1 + \beta x_k} - \mu\|^2_{\mathcal{H}}  = \frac{\|\hat{\mu} - \mu\|^2_{\mathcal{H}}}{(1 + \beta x_k)^2}$.
For strategy pair $(x, x')$, the difference in distances is:
   $\frac{\|\hat{\mu} - \mu\|^2_{\mathcal{H}}}{(1 + \beta x_k)^2} - \frac{\|\hat{\mu} - \mu\|^2_{\mathcal{H}}}{(1 + \beta x'_k)^2} = \|\hat{\mu} - \mu\|^2_{\mathcal{H}}(\frac{1}{(1 + \beta x_k)^2} - \frac{1}{(1 + \beta x'_k)^2})$.

Since $x_k > x'_k$:
   $\frac{1}{(1 + \beta x_k)^2} - \frac{1}{(1 + \beta x'_k)^2} < 0$.
Let $A = \|\hat{\mu} - \mu_k\|^2_{\mathcal{H}}$ and $B = \|\hat{\mu} - \mu'_k\|^2_{\mathcal{H}}$. The theorem's condition becomes:
   $A(\frac{1}{(1 + \beta x_k)^2} - \frac{1}{(1 + \beta x'_k)^2}) \leq B(\frac{1}{(1 + \beta x_k)^2} - \frac{1}{(1 + \beta x'_k)^2})$.

Given the negative common factor from step 4, this inequality holds if and only if:
   $\|\hat{\mu} - \mu_k\|^2_{\mathcal{H}} \geq \|\hat{\mu} - \mu'_k\|^2_{\mathcal{H}}$.
\end{proof}

\section{Proof for \cref{lemma:concavity}}\label{proof:lemma:concavity}

\textbf{Lemma \ref{lemma:concavity}} (Concavity of Client Utility Function).
Given fixed strategies $\boldsymbol{x}_{-i}$ of other clients, the utility function $U_i$ of client~$i$ is concave in $x_i$ for $x_i \in [0, 1]$.
\begin{proof}
    First, let's examine the second derivative of $u_i = \hat{\lambda}_q \|\mu(x) - \mu_k\|^2_{\mathcal{H}}$ with respect to $x_i$:
   \(\frac{\partial^2}{\partial x_k^2} \|\mu(x) - \mu_k\|^2_{\mathcal{H}} = \frac{6\alpha_k^2}{S^2}\|\mu(x) - \mu_k\|^2_{\mathcal{H}} \geq 0\)

Then, recalling client utility function~\eqref{eq:client_opt}, we derive the second derivative of $U_i$: $\frac{\partial^2 U_i(\boldsymbol{x} )}{\partial x_i^2} \leq 0$. 
\end{proof}

\section{Proof for \cref{theorem:nash_existence}}\label{proof:theorem:nash_existence}

\textbf{Proposition \ref{theorem:nash_existence}} (Existence of Nash Equilibrium).
Given the server's payment scheme $\boldsymbol{p} = (p_1, \dots, p_N)$, there exists a Nash equilibrium in the client-side game. 

\begin{proof} To prove the existence of a Nash equilibrium, we will use the Debreu-Glicksberg-Fan theorem.
    First, the strategy space $[0,1]^N$ is non-empty, compact, and convex, and $U_i$ is continuous in $\boldsymbol{x}$.
     By \cref{lemma:concavity}, $U_i$ is concave in $x_i$. 
    Then, let's define the best response correspondence for each client~$i$:
    \(
    BR_i(\boldsymbol{x}_{-i}) = \text{argmax}_{x_i \in [0,1]} U_i(x_i, \boldsymbol{x}_{-i}; p_i)
    \).
    Therefore, there exists fixed $\boldsymbol{x}^* \in BR(\boldsymbol{x}^*)$, which is a Nash equilibrium.
\end{proof}

\section{Proof for \cref{theorem:nash_characterization}}\label{proof:theorem:nash_characterization}

\textbf{Theorem \ref{theorem:nash_characterization}} (Characterization of Nash Equilibrium).
In any Nash Equilibrium $\boldsymbol{x}^*$, for each client $k \in \mathcal{N}$, given payment rate $p_k$, the equilibrium strategy $x_k^*$ is characterized as follows:
\begin{equation}
x_k^*(p_k) = \begin{cases}
0 & \text{if } p_k < \underline{p_k}  \\
\left( \frac{\phi(\boldsymbol{x}^*_{-k})}{\alpha_k^2 (c_k-p_k)} \right)^{1/3}  - \frac{S_0}{\alpha_k} & \text{if } \underline{p_k} \leq p_k \leq \overline{p_k} \\
1 & \text{if } p_k > \overline{p_k}
\end{cases}
\end{equation}
where $\phi(\boldsymbol{x}^*_{-k}) = 2 \hat{\lambda}_q\| \sum_{i \neq k} \alpha_i x^*_i (\mu_i - \mu_k)\|_\mathcal{H}^2$, and the threshold payment are:
\begin{equation}
\underline{p_k} = c_k- \frac{\alpha_k \phi(\boldsymbol{x}_{-k}^*)}{S_0^{3}}, \quad \overline{p_k} = c_k - \frac{\alpha_k \phi(\boldsymbol{x}_{-k}^*)}{(S_0 + \alpha_k)^3} \,.
\end{equation}

\begin{proof} By the concavity of $U_i$ (\cref{lemma:concavity}),  
we proceed by analyzing the first-order condition of the client's utility maximization problem.

Taking the derivative of the utility function of client~$i$ with respect to $x_i$: $ \frac{\partial U_k}{\partial x_k} 
= p_k + \frac{2\alpha_k \hat{\lambda}_q}{(S_0 + \alpha_k x_k)^3}\| \sum_{i \neq k} \alpha_i x_i (\mu_i - \mu_k)\|_{\mathcal{H}}^2 - c_k$,
% \begin{align*}
%     \frac{\partial U_k}{\partial x_k} & = p_k + \frac{2\alpha_k \hat{\lambda}_q}{S}\|\mu(x) - \mu_k\|^2_{\mathcal{H}} - c_k \\
%     & = p_k + \frac{2\alpha_k \hat{\lambda}_q}{(S_0 + \alpha_k x_k)^3}\| \sum_{i \neq k} \alpha_i x_i (\mu_i - \mu_k)\|_{\mathcal{H}}^2 - c_k 
% \end{align*}
where $S = \sum_i \alpha_i x_i =  S_0 + \alpha_k x_k$.

Solving $\frac{\partial U_k}{\partial x_k} = 0$, we obtain:
$
\tilde{x}_k(p_k) = \left( \frac{\phi(\boldsymbol{x}_{-k})}{\alpha_k^2 (c_k-p_k)} \right)^{1/3}  - \frac{S_0}{\alpha_k}
$,
where $\phi(\boldsymbol{x}_{-k}) = 2 \hat{\lambda}_q\| \sum_{i \neq k} \alpha_i x_i (\mu_i - \mu_k)\|_{\mathcal{H}}^2$.

The threshold values $\underline{p_i}$ and $\overline{p_i}$ is derived in \eqref{eq:threshold_p}.
\end{proof}

\section{Proof for \cref{theorem:suff_uniqueness_nash}}\label{proof:theorem:suff_uniqueness_nash}

\textbf{Theorem \ref{theorem:suff_uniqueness_nash}} (Sufficient Condition for Uniqueness of Nash Equilibrium).
For the Nash Equilibrium (NE) characterized in Theorem~\ref{theorem:nash_characterization}, if for  $\forall k \in \mathcal{N}$:
$
D < \frac{3 \alpha_k (1-\alpha_k)^2}{4}  \sqrt{ \frac{3 |c_k-p_k|}{\hat{\lambda}_q}} \,
$, 
where $D =\max_{i, j \in \mathcal{N}} \|\mu_i - \mu_j\|_{\mathcal{H}}$.
Then the NE is unique.

\begin{proof}
First, we analyze the derivative of the best response function. 
Without loss of generality, we consider the interior case where $\underline{p_k}(\boldsymbol{x}_{-k}) \leq p_k \leq \overline{p_k}(\boldsymbol{x}_{-k})$, we derive:
\begin{equation*}
\frac{\partial BR_k}{\partial x_j} = \frac{1}{3} \left( \frac{\phi(\boldsymbol{x}_{-k})}{\alpha_k^2 (c_k-p_k)} \right)^{-2/3} \frac{\partial \phi(\boldsymbol{x}_{-k})}{\partial x_j} \frac{1}{\alpha_k^2 (c_k-p_k)}
\end{equation*}
and $\frac{\partial \phi(\boldsymbol{x}_{-k})}{\partial x_j} = 4\hat{\lambda}_q\alpha_j \langle \mu_j - \mu_k, \sum_{l \neq k} \alpha_l x_l (\mu_l - \mu_k) \rangle_{\mathcal{H}}$.

Next, 
to show contraction, we need to bound $\sum_{j \neq k} |\frac{\partial BR_k}{\partial x_j}|$: 
Let $D = \max_{i, j} \|\mu_i - \mu_j\|_{\mathcal{H}}$, then

$|\frac{\partial \phi(\boldsymbol{x}_{-k})}{\partial x_j}| \leq 4 \hat{\lambda}_q D^2 \sum_{j \neq k} \alpha_j \sum_{l \neq k} \alpha_l x_l$

Since $x_l \in [0,1]$ for all $l$: $|\frac{\partial \phi(\boldsymbol{x}_{-k})}{\partial x_j}| \leq 4 \hat{\lambda}_q D^2$, we have
$
\sum_{j \neq k} |\frac{\partial BR_k}{\partial x_j}| \!\leq\! \frac{4 \hat{\lambda}_q D^2 (\sum_{j \neq k} \alpha_j)^2}{3\alpha_k^2 (c_k-p_k)} \left( \frac{\phi(\boldsymbol{x}_{})}{\alpha_k^2 (c_k-p_k)} \right)^{-2/3}
\,.$

Consequently,
for $BR_k$ to be a contraction, we need:
$\frac{4 \hat{\lambda}_q D^2 }{3\alpha_k^2 (c_k-p_k)} \left( \frac{\phi(\boldsymbol{x}_{-k})}{\alpha_k^2 (c_k-p_k)} \right)^{-2/3} < 1$.

Recall $\phi(\boldsymbol{x}_{-k})$ defined in \cref{theorem:nash_characterization}, 
$BR_k$ is a contraction under:
$
\max_{i, j} \|\mu_i - \mu_j\|_{\mathcal{H}} < \frac{3 \alpha_k m_k^2}{4}  \sqrt{ \frac{3 |c_k-p_k|}{\hat{\lambda}_q}}
$.

Note that $m_k = \sum_{j \neq {k}} \alpha_j x_j \leq 1 - \alpha_k$, then for all clients: 
\begin{equation}\label{eq:nash_unique_sufficient_1}
\max_{i, j} \|\mu_i - \mu_j\|_{\mathcal{H}} < \frac{3 \alpha_k (1-\alpha_k)^2}{4}  \sqrt{ \frac{3 |c_k-p_k|}{\hat{\lambda}_q}}
\end{equation}

Thus, if the condition \eqref{eq:nash_unique_sufficient_1} holds for all clients $k$, then the joint best response function $BR(\boldsymbol{x}) = \{BR_k(\boldsymbol{x}_{-k})\}_{k \in \mathcal{N}}$ is a contraction mapping.
By the Banach fixed-point theorem~\cite{van2000existence}, under~\eqref{eq:nash_unique_sufficient_1}, the Nash Equilibrium is unique.
\end{proof}

\section{Proof for \cref{theorem:quasiconvex_Us}}\label{proof:theorem:quasiconvex_Us}

\textbf{Lemma \ref{theorem:quasiconvex_Us}} (Quasiconvexity of $\tilde{u}_{server}$).
The function $\tilde{u}_{server}(\boldsymbol{x}^*(\boldsymbol{p}), \boldsymbol{p})$ is quasi-convex.

\begin{proof}

First, $\tilde{u}_{server}$ is clearly monotonic in $\mu(\boldsymbol{x}^*(\boldsymbol{p}))$. Define $h(\boldsymbol{p}) = \mu(\boldsymbol{x}(\boldsymbol{p}))$. We will show $h(\boldsymbol{p})$ is quasilinear. The function 
$ h(\boldsymbol{p}) = \frac{\sum_{j \in \mathcal{N}} \alpha_jx_j \mu_j}{\sum_{j \in \mathcal{N}} \alpha_jx_j}
$ is linear-fractional, thus quasilinear.
% From Theorem \ref{theorem:nash_characterization}, we know $\frac{\partial x_i}{\partial p_i} \geq 0$, so $x_i$ is non-decreasing in $p_i$. The composition of a quasilinear function with a non-decreasing function is quasilinear. Thus, $h(\boldsymbol{p})$ is quasilinear.
% Note that $\tilde{u}_{server}$ is monotonic in $\mu(\mathbf{x^*}(\boldsymbol{p}))$, which we've shown to be quasilinear with respect to $\boldsymbol{p}$. Therefore, $\tilde{u}_{server}(\mu(\mathbf{x^*}(\boldsymbol{p})))$ is quasiconvex with respect to $\boldsymbol{p}$. 
From Theorem \ref{theorem:nash_characterization}, $\frac{\partial x_i}{\partial p_i} \geq 0$, so $x_i$ is non-decreasing in $p_i$. The composition of a quasilinear function with a non-decreasing function is quasilinear. Hence, $h(\boldsymbol{p})$ is quasilinear. Consequently, $\tilde{u}_{server}(\mu(\mathbf{x}^*(\boldsymbol{p})))$ is quasiconvex with respect to $\boldsymbol{p}$.
\end{proof}

\section{Proof for \cref{lemma:monotonicity_px}}
\label{app:mono-px}

\textbf{Lemma \ref{lemma:monotonicity_px}}.
The function $\boldsymbol{p}^{\top} \boldsymbol{x}^*(\boldsymbol{p})$ is monotonically decreasing in $\boldsymbol{p}$ for $p_i \in [\underline{p_i}, \overline{p_i}]$, $\forall i \in \mathcal{N}$.

\begin{proof}
    Let $h(\boldsymbol{p}) \!=\!\boldsymbol{p}^{\top}\! \boldsymbol{x}^*(\boldsymbol{p}) \!=\! \sum_{i \in \mathcal{N}} \!\left( \left( \frac{p_i^3 \phi(\boldsymbol{x}^*_{-i})}{\alpha_i^2 (c_i-p_i)} \right)^{1/3}  \!-\! \frac{S_0 p_i}{\alpha_i}\!\right)$ for $p_i \in [\underline{p_i}, \overline{p_i}]$.
    Define $g(p_i) = p_i^3 \frac{\phi(\boldsymbol{x}_{-i}^*)}{\alpha_i^2 (c_i-p_i)}$, and we can show $g(p_i)$ is monotonically decreasing, thus $h(\boldsymbol{p})$ is monotonically decreasing for $p_i \in [\underline{p_i}, \overline{p_i}]$.\looseness=-1
\end{proof}

\end{document}